\documentclass[journal]{IEEEtran}
\usepackage{amssymb}
\usepackage[cmex10]{amsmath}
\usepackage{stfloats}
\usepackage{graphicx}
\usepackage{subfigure}
\usepackage{tabularx}
\usepackage{epsfig,epsf,color,balance,cite}
\usepackage{verbatim}
\usepackage{url}
\usepackage{bm}

\newtheorem{theorem}{Theorem}
\newtheorem{coro}{Corollary}
\usepackage{algorithm}
\usepackage{algpseudocode}
\begin{document}

\title{Joint Time Allocation and Power Control in Multi-Cell Networks with Load Coupling: Energy Saving and Rate Improvement}

\author{
\IEEEauthorblockN{Zhaohui Yang, \IEEEmembership{Student Member, IEEE},
                  Cunhua Pan, \IEEEmembership{Member, IEEE},
                  Wei Xu, \IEEEmembership{Senior Member, IEEE},
                  Hao Xu, \IEEEmembership{Student Member, IEEE} and
                  Ming Chen \IEEEmembership{Member, IEEE}
                  }
\thanks{Z. Yang, W. Xu, H. Xu and M. Chen are with the National Mobile Communications Research
Laboratory, Southeast University, Nanjing 210096, China.  (Email: yangzhaohui@seu.edu.cn, wxu.seu@gmail.com, xuhao2013@seu.edu.cn, chenming@seu.edu.cn).}
 \thanks{C. Pan is with the School of Electronic Engineering and Computer Science, Queen Mary, University of London, London E1 4NS, UK (Email: c.pan@qmul.ac.uk).}
}
\maketitle

\IEEEpeerreviewmaketitle

\begin{abstract}
In this paper, we consider the problems of minimizing sum power and maximizing sum rate for multi-cell networks with load coupling, where coupling relation occurs among cells due to inter-cell interference. This coupling relation is characterized by the signal-to-interference-and-noise-ratio (SINR) coupling model with cell load vector and cell power vector as the variables. Due to the nonlinear SINR coupling model, the optimization problems for multi-cell networks with load coupling is nonconvex. To solve these nonconvex problems, we first consider the optimization problems for single-cell networks. Through variable transformations, the optimization problems can be equivalently transformed into convex problems. By solving the Karush-Kuhn-Tucker (KKT), the optimal solutions to power minimization and rate maximization problems can be obtained in closed form. Based on the theoretical findings of optimization problems for single-cell networks, we develop a distributed time allocation and power control algorithm with low complexity for sum power minimization in multi-cell networks. This algorithm is proved to be convergent and globally optimal by using the properties of standard interference function. For sum rate optimization in multi-cell networks, we also provide a distributed algorithm which yields suboptimal solution. Besides, the convergence for this distributed algorithm is proved. Numerical results illustrate the theoretical findings, showing the superiority of our solutions compared to the conventional solution of allocating uniform power for users in the same cell.
\end{abstract}

\IEEEpeerreviewmaketitle

\section{Introduction}
Sum power minimization, and sum rate maximization are two fundamental optimization problems in wireless communication networks.
To solve these two problems, resource allocation and power control are often considered \cite{793310,1244793,1545851,6936359,6815733,4289518,Georgiadis2006Resource,6826535}.

In time-division multiple access (TDMA) networks, the base station (BS) serves multiple users through time division.
Since users in the same cell can be allocated with different resources by time division, intra-cell interference among users is eliminated.
Due to this distinction, wireless powered communications networks using TDMA has arisen a great interest \cite{6678102,7328715,7098429,7037291}.
The above works all assumed a single-cell network.
In practical networks, there are multiple cells and different cells share the same resources due to the limitation of resources \cite{6824752,Pan2015Totally,Pan2016Pricing}.
Hence, inter-cell interference among different cells is inevitable, resulting that the optimization algorithm for a single-cell network cannot be directly applied to a multi-cell network.
Thus, the optimization problem of a multi-cell network with TDMA is of great importance.

To characterize the inter-cell interference among cells, the load coupling model was proposed in \cite{Siomin2012Analysis} for a multi-cell network with OFDM.
In \cite{Siomin2012Analysis}, the considered OFDM network assumed frequency-flat fading channels and only time resources were scheduled.
Thus, the so-called OFDM~\footnote{In this paper, the OFDM network is the same as TDMA network since only time resources are scheduled.} network in \cite{Siomin2012Analysis} is equivalent to a TDMA network.
Different from the conventional resource allocation problems \cite{4027580,6294504,4278409,5773453,4489657,4698488}, where the subchannel assignment problem with integer variable is always considered, the variables in resource allocation problems \cite{fehske2012aggregation,siomina2014constrained,ho2014data,siomina2012load,siomina2013optimization,DBLPYouY16} with load are always continuous.
The load of a cell in \cite{Siomin2012Analysis} is defined as the average level of usage of time resources.
Specifically, the load coupling model shows that the high load of a BS means a high probability for other BSs to receive interference.
The load coupling model with fixed power has been shown to give a good approximation for a multi-cell network especially at high data arrival rates in \cite{fehske2012aggregation}.
Since the load coupling model has a good structure with high accuracy for characterizing inter-cell interference, it has been used in many applications \cite{siomina2014constrained}, such as data offloading \cite{ho2014data}, load balancing \cite{siomina2012load}, location planning \cite{siomina2013optimization}, and user association \cite{DBLPYouY16} in multi-cell networks.

Previous works \cite{Siomin2012Analysis,fehske2012aggregation,siomina2014constrained,ho2014data,siomina2012load,siomina2013optimization,DBLPYouY16} using load coupling model all assumed fixed transmission power of BSs.
To tackle the problem of minimizing the sum transmission power in multi-cell networks with load coupling, both load and power are optimization variables. 
Recently, \cite{Chin2015Power} considered the sum power minimization problem where both load and power of each cell are incorporated into the signal-to-interference-noise-ratio (SINR) coupling model.
The coupling was implicitly characterized with load and power as the variables of interest using non-linear load and power coupling equation.
It was analytically shown that operating at full load is optimal, and an iterative power adjustment algorithm for all BSs was provided to achieve the full load.
In \cite{Chin2015Power}, the transmission power is different for users in different cells, but the same for users in the same cell.
However, exploiting multiuser diversity, the sum power consumption can be additionally improved by optimizing the power allocation in the same cell.
Moreover, the sum rate maximization problem for multi-cell networks with load coupling is rarely considered.

In this paper, we consider multiuser diversity by allocating users in the same cell with unequal power and fractions of resources.
We aim to jointly optimize time allocation and power control for sum power minimization and sum rate maximization in multi-cell networks with load coupling.
Since the power vectors in the Lagrangian of sum power minimization and sum rate maximization cannot be decoupled, it is difficult to establish the distributed algorithm using alternating directions method of multipliers (ADMM) \cite{6484993,6156468,6748974}.
Instead of using ADMM, we propose two distributed algorithms with low complexity and limited information exchange.
Based on the optimization techniques, the main contributions in this paper are summarized as follows:

\begin{enumerate}
  \item By using variable transformations, sum power minimization and sum rate maximization problems for a single-cell network with load coupling can be transformed into equivalent convex problems.
      We provide the globally optimal solutions in closed form to these two optimization problems by solving the Karush-Kuhn-Tucker (KKT) conditions.
      Using the KKT conditions, we prove that operating with full time resources is optimal to minimize sum power and maximize sum rate.
      Besides, we also give the optimal conditions for power control of these two problems.
  \item To minimize sum power of BSs for a multi-cell network with load coupling, we extend the load and power coupling model in \cite{Chin2015Power} for the case where users in the same cell are allocated with unequal power.
      We develop a distributed time allocation and power control algorithm 
      for sum power minimization.
      The distributed algorithm is provided along with its convergence and global optimality proof by using the properties of standard interference function.
  \item Based on the optimization problems in a single-cell network with load coupling, we propose a distributed time allocation and power control algorithm for sum rate maximization in a multi-cell network with load coupling.
      We also prove the convergence of this distributed algorithm.
\end{enumerate}

This paper is organized as follows.
In Section $\text{\uppercase\expandafter{\romannumeral2}}$, we introduce the system model.
Section $\text{\uppercase\expandafter{\romannumeral3}}$ gives two common optimization problems for single-cell networks and provides the globally optimal solutions to these two problems.
Sum power minimization and sum rate maximization for multi-cell networks are addressed in Section $\text{\uppercase\expandafter{\romannumeral 4}}$ and Section $\text{\uppercase\expandafter{\romannumeral 5}}$, respectively.
Some numerical results are displayed in Section $\text{\uppercase\expandafter{\romannumeral6}}$
and conclusions are finally drawn in Section $\text{\uppercase\expandafter{\romannumeral7}}$.

\section{Joint Time Allocation and Power Control Problems for Multi-Cell Networks}
Consider a multi-cell network with load coupling, where the set of BSs is denoted by ${\cal{N}}=\{1, 2, \cdots, N\}$.
Each BS $i \in {\cal{N}}$ serves one unique group of users, denoted by set
${\cal{J}}_i=\{J_{i-1}+1,J_{i-1}+2,\cdots, J_{i}\}$, where $J_0=0$, $J_i=\sum_{l=1}^i|{\cal{J}}_l|$, $|\cdot|$ is the cardinality of a set and $|{\cal{J}}_i|\geq 1$.
We focus on the
downlink communication scenarios where mutual interference exists among cells.
Denote the power spectral density of BS $i$ for user $j \in {\cal{J}}_i$ by $p_{ij}$ \cite{Siomin2012Analysis,siomina2013optimization,Chin2015Power}.
For notational convenience, we collect all power as vector
$\pmb{p}=(p_{11}, \cdots, p_{1J_1}, \cdots, p_{N(J_{N-1}+1)}, \cdots, p_{NJ_N})^T$.
User $j \in {\cal{J}}_i$ is served by BS $i$ at achievable rate $r_{ij}$ that has to be greater than a rate demand $D_{ij}>0$.
We collect all the rates as vector $\pmb{r}=(r_{11}, \cdots, r_{1J_1}, \cdots, r_{N(J_{N-1}+1)}, \cdots, r_{NJ_N})^T$ and the corresponding minimal rate demands as vector $\pmb{D}=(D_{11}, \cdots, D_{1J_1}, \cdots, D_{N(J_{N-1}+1)}, \cdots, D_{NJ_N})^T$.
Thus, the rate vector meets the rate demand constraints if $\pmb{r} \geq \pmb{D}$.

In this paper, we consider the load and power coupling model as in \cite{Siomin2012Analysis}.
Denote $m_{ij}$ as the fraction of resources that are allocated to user $j \in {\cal{J}}_i$ in cell $i$ by time division.
Collect all the time fraction factors as vector $\pmb{m}=(m_{11}, \cdots, m_{1J_1}, \cdots, m_{N(J_{N-1}+1)}, \cdots, m_{N J_N})^T$, which can be viewed as load vector in \cite{Siomin2012Analysis}.
The load of BS $i$ can be calculated by the summation of load for serving every user $j \in {\cal{J}}_i$, which should satisfy $\sum_{j \in {\cal{J}}_i }m_{ij} \leq 1$.
It can be observed that the average power of BS $i$ is $\sum_{j \in {\cal{J}}_i }m_{ij} p_{ij}$.
The average power of each BS always has a maximal power limit, i.e., $\sum_{j \in {\cal{J}}_i }m_{ij} p_{ij} \leq P_i^{\max}$.
If the time allocation is randomly distributed and we consider the long-term average interference from other BSs,
the average SINR of user $j$ associated with BS $i$ can be expressed as \cite{Siomin2012Analysis,fehske2012aggregation,siomina2014constrained,ho2014data,siomina2012load,siomina2013optimization,DBLPYouY16}
\begin{equation}\label{eq1}
\eta_{ij}=\frac{p_{ij} g_{ij}} {\sum_{k \in {\cal{N}}\setminus \{i\}} \sum_{l \in {\cal{J}}_k} m_{kl} p_{kl} g_{kj} +\sigma ^2},
\end{equation}
where $g_{ij}$ is the channel gain from BS $i$ to user $j$, $\mathcal A\setminus \mathcal B=\{k|k\in\mathcal A, k \notin \mathcal B\}$, and $\sigma^2$ represents the noise power density.
Intuitively, $m_{kl}$ can be interpreted as the probability of receiving interference from BS $k$ for serving $l \in \mathcal J_k$ in a long time (see Fig. 1).
Thus, the combined term $m_{kl} p_{kl} g_{kj}\in [0, p_{kl} g_{kj}]$ is interpreted as the average interference taken over time.
Equation (\ref{eq1}) with averaged interference power evaluated by load variables has been shown to give a good approximation for a multi-cell network especially at high data arrival rates \cite{fehske2012aggregation}.
Thus, Equation (\ref{eq1}) has been used in many applications \cite{siomina2014constrained,ho2014data,siomina2012load,siomina2013optimization,DBLPYouY16},
as it has a good structure with high accuracy for characterizing inter-cell interference.

The achievable rate $r_{ij}$ of user $j \in {\cal{J}}_i$ can be formulated as
\begin{eqnarray}\label{eq2}
r_{ij}&&\!\!\!\!\!\!\!\!\!\!
=B m_{ij}\log_2\left(1+ \frac{p_{ij} g_{ij}} {\sum_{k \in {\cal{N}}\setminus \{i\}} \sum_{l \in {\cal{J}}_k} m_{kl} p_{kl} g_{kj} +\sigma ^2} \right)
\nonumber\\
&&\!\!\!\!\!\!\!\!\!\!
\triangleq f_{ij}(\pmb{m},\pmb{p}), \quad\forall i  \in {\cal{N}}, \forall j \in {\cal{J}}_i,
\end{eqnarray}
where $B$ is the system bandwidth.
From nonlinear Equation (\ref{eq2}), it is observed that the rate of a user in a specific cell is coupled with the load and power of other cells.

\begin{figure}
\centering
\includegraphics[width=2.4in]{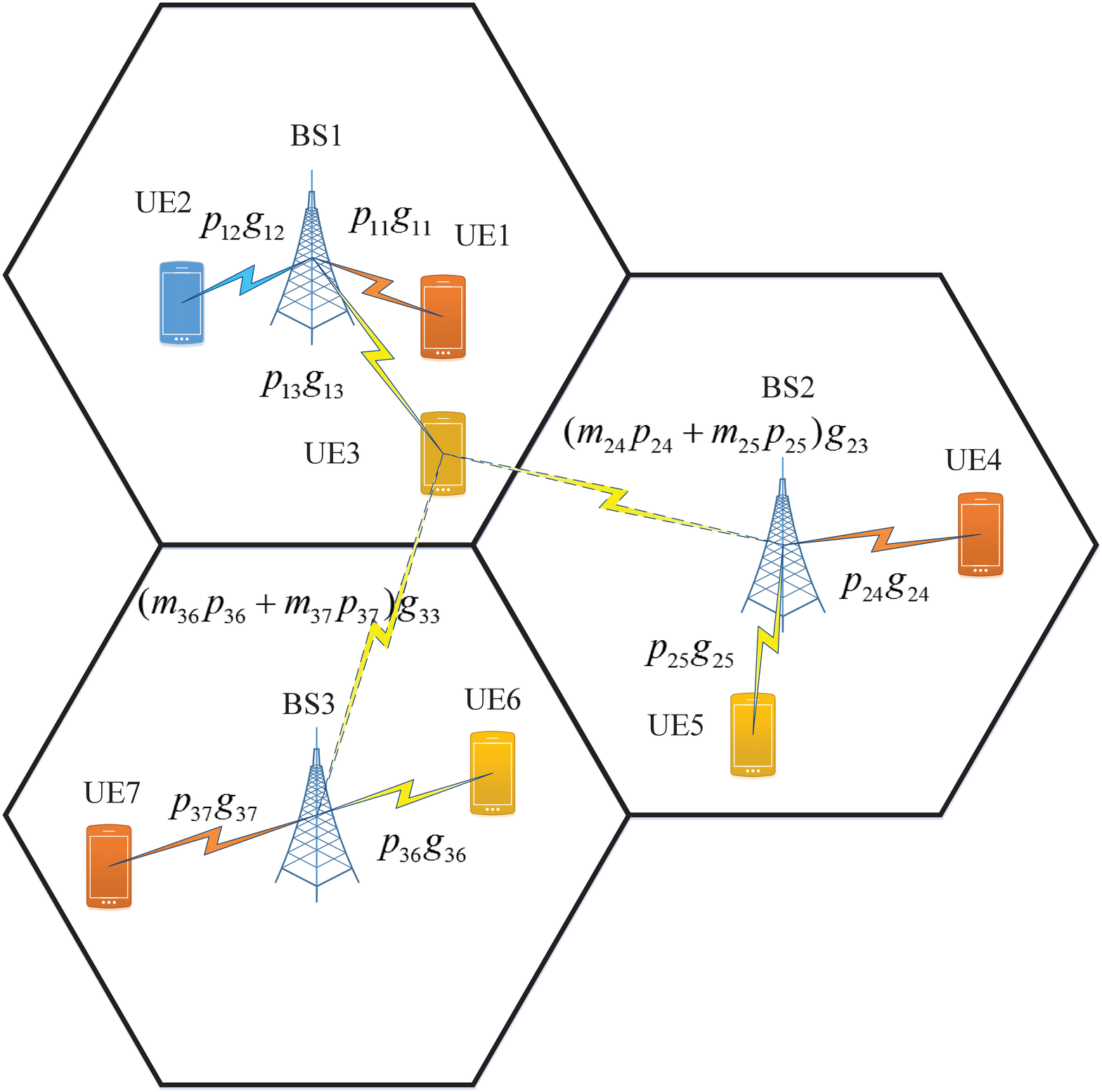}
\vspace{-1em}
\caption{System model for multi-cell networks with load coupling.  \label{fig0}}
\vspace{-1em}
\end{figure}

We aim at solving the sum power and rate optimization problems, subject to the power and load constraints for each BS and the rate demand constraint for every user.
Formally, the time allocation and power control problem can be formulated as
\begin{subequations}\label{max1_1}
\begin{align}
\mathop{\min}_{
\pmb{0} \leq \pmb{m}, \pmb{0}\leq \pmb{p}, \pmb 0 \leq \pmb{r}} \quad
&V(\pmb m, \pmb p, \pmb r)
\\
\textrm{s.t.}\qquad\quad
&r_{ij}=f_{ij}(\pmb{m},\pmb{p}), \quad\forall i  \in {\cal{N}}, \forall j \in {\cal{J}}_i\\
&\sum_{j \in {\cal{J}}_i }m_{ij} p_{ij} \leq P_i^{\max}, \quad\forall i  \in {\cal{N}}\\
&\sum_{j \in {\cal{J}}_i }m_{ij} \leq 1, \quad  \forall i  \in {\cal{N}}\\
&\pmb D  \leq  \pmb r,
\end{align}
\end{subequations}
where $V(\pmb m, \pmb p, \pmb r)$ is the objective function, which can be sum power $\sum_{i \in \mathcal N} \sum_{j\in\mathcal J_i} m_{ij}p_{ij}$ and negative sum rate $-\sum_{i \in \mathcal N} \sum_{j\in\mathcal J_i} r_{ij}$.

Obviously, the feasible set of Problem (\ref{max1_1}) is nonconvex.
Obtaining global optimum of time allocation and power control problems for multi-cell networks is known to be difficult even by the centralized algorithm.
In the following, we first solve the time allocation and power control problems for single-cell networks.
Based on the theoretical findings for single-cell networks,
we devise two novel distributed algorithms to deal with time allocation and power control problems for multi-cell networks with low computational complexity, respectively.
Interestingly, the distributed algorithm for sum power minimization in multi-cell networks can be proved to be globally optimal. 

\section{Joint Time Allocation and Power Control for Single-Cell Networks}
In this section, we consider a downlink single-cell network with $M$ users.
Denote the time division factor of user $j$ by $m_j$, which should satisfy
\begin{equation}\label{1RBs_T_Constraint}
\sum_{j=1}^{M}m_j \leq 1.
\end{equation}
In a single-cell network without inter-cell interference, the achievable rate $r_j$ of user $j$ can be formulated as
\begin{equation}\label{2Bits_of_user}
r_j=B m_j \log_2\left(1+\frac{p_j g_j}{\sigma^2}\right),
\end{equation}
where $p_j$ is the power spectral density of the BS for user $j$, and $g_j$ denotes the channel gain between the BS and user $j$. 
The average transmission power $p$ of the BS can be calculated by summing the power for serving all users,
\begin{equation}\label{4Average_power_of_BS}
p=\sum_{j=1}^M m_j p_j.
\end{equation}

Now, it is ready to formulate the sum power and rate optimization problems for a single-cell network as
\begin{subequations}\label{OptimizationSingleCell_1}
\begin{align}
\mathop{\min}_{\pmb 0\leq \pmb{\tilde m}, \pmb 0 \leq \pmb{\tilde p}} \quad\!\!\!\!
&U({\pmb{\tilde m}, \pmb{\tilde p}})\\
\textrm{s.t.}\qquad   \!\!\!\!
& D_j \!\leq\!B m_j \log_2\left(\!1\!+\!\frac{g_j  p_j}{\sigma^2 }\!\right)\!,  \quad j \!=\!1, \cdots, M\\
&\sum_{j=1}^{M}m_j \leq 1\\
& {\sum_{j=1}^M m_j p_j} \leq P_{\max},
\end{align}
\end{subequations}
where $\pmb{\tilde m}=(m_1, \cdots, m_M)^T$, $ \pmb{\tilde p}=( p_1, \cdots, p_M)^T$, $D_j>0$ is the minimal transmission rate of user $j$,
and $P_{\max}$ is the maximal power of the BS.
$U({\pmb{\tilde m}, \pmb{\tilde p}})$ is the objective function, which can be sum power $\sum_{j=1}^M p_j$ or negative sum rate $-\sum_{j=1}^M B m_j \log_2\left(1+\frac{g_j  p_j}{\sigma^2 }\right)$.

Obviously, the feasible set of Problem (\ref{OptimizationSingleCell_1}) is nonconvex due to constraints (\ref{OptimizationSingleCell_1}b) and (\ref{OptimizationSingleCell_1}d).
In this case,
we introduce a set of new variables $\bar p_j= m_j p_j$, $j=1, \cdots, M$.
The new variable $\bar p_j$ can be viewed as the average transmission power of the BS for user $j$ in the scheduling time.
Denoting $ \pmb{\bar p}=(\bar p_1, \cdots, \bar p_M)^T$, Problem (\ref{OptimizationSingleCell_1}) is equivalent to the following problem:
\begin{subequations}\label{OptimizationSingleCell}
\begin{align}
\mathop{\min}_{\pmb 0\leq \pmb{\tilde m}, \pmb 0 \leq \pmb{\bar p}} \quad\!\!\!\!
&\bar U({\pmb{\tilde m}, \pmb{\bar p}})\\
\textrm{s.t.}\qquad   \!\!\!\!\!
& D_j \!\leq\!B m_j \log_2\left(\!1\!+\!\frac{g_j \bar p_j}{\sigma^2 m_j}\!\right)\!,  \!\!\quad j \!=\!1, \cdots, M\\
&\sum_{j=1}^{M}m_j \leq 1\\
& {\sum_{j=1}^M \bar p_j} \leq P_{\max},
\end{align}
\end{subequations}
where we define $B m_j \log_2\left(1+\frac{g_j \bar p_j}{\sigma^2 m_j}\right)=0$ for $m_j=0$.
It can be easily verified that the feasible set of Problem (\ref{OptimizationSingleCell}) is convex.
In the following, we solve the sum power minimization problem, and sum rate maximization problem, separately.
\subsection{Sum Power Minimization Problem}
We consider the sum power minimization problem with objective function
\begin{equation}
\bar U({\pmb{\tilde m}, \pmb{\bar p}})={\sum_{j=1}^M \bar p_j},
\end{equation}
and constraints (\ref{OptimizationSingleCell}b)-(\ref{OptimizationSingleCell}c).
The objective function is linear and the sum power minimization problem is convex.
The optimal solution of Problem (\ref{OptimizationSingleCell}) can be effectively obtained in closed form by solving the KKT conditions as in Appendix A.

\subsection{Sum Rate Maximization Problem}
We consider the sum rate maximization problem with objective function
\begin{equation}
\bar U({\pmb{\tilde m}, \pmb{\bar p}})=-{\sum_{j=1}^M  B m_j \log_2\left(1+\frac{g_j \bar p_j}{\sigma^2 m_j}\right)},
\end{equation}
and constraints (\ref{OptimizationSingleCell}b)-(\ref{OptimizationSingleCell}d).
Obviously, the sum rate maximization problem is convex.
To obtain the globally optimal solution in closed form,
the details of solving KKT conditions can be found in Appendix B.

From Appendix A and Appendix B, we have the following theorem about the optimal conditions.
\begin{theorem}
To minimize sum power and maximize sum rate, it is optimal to transmit with full time resources, i.e., $\sum_{j=1}^M m_{j}^*=1$.
To minimize sum power, the time and power are optimized to just satisfy the minimal rate requirements.
To maximize sum rate, it is optimal to transmit with maximal power, i.e., $\sum_{j=1}^M \bar p_{j}^*=P_{\max}$.
Besides, the optimal power allocation strategy is to allocate
the additional power to the user with the best channel gain,
while other users are allocated with the minimal power to
maintain their minimal rate requirements.
\end{theorem}

Theorem 1 is different from the conventional waterfilling algorithm \cite[Section~5.3.3]{tse2006fundamentals}.
According to the waterfilling algorithm, the transmitter allocates more power to the strong users, taking advantage of the better channel conditions, and less or even no power to the weaker ones.
This means that the additional power can be allocated to some users with better channel conditions in the waterfilling algorithm, while the proposed sum rate maximization algorithm for single-cell networks states that the additional power should be allocated to only one user with the strongest channel gain.
The reason is that the proposed  sum rate maximization Problem (8) considers joint time allocation and power control, while only power control is considered in the conventional power control problem \cite[Section~5.3.3]{tse2006fundamentals}.

\section{Sum Power Minimization Problem for Multi-Cell Networks}
\label{section3}
In this section, based on the results about sum power minimization for single-cell networks, we solve the sum power minimization Problem (\ref{max1_1}) with $V(\pmb m, \pmb p, \pmb r)=\sum_{i \in \mathcal N} \sum_{j\in\mathcal J_i}$ $m_{ij}p_{ij}$.
We establish a distributed time allocation and power control algorithm, and provide the proof of convergence and global optimality for this distributed algorithm.

\subsection{Optimal Condition}
We establish the optimal condition for rate vector $\pmb r$ and load vector $\pmb m$ as follows.
\begin{theorem}\label{theoremopcmp}
If Problem (\ref{max1_1}) is feasible, the optimal solution to sum power minimization problem is such that the rate vector reaches the minimal rate constraints $\pmb r^*=\pmb D$, and the load vector satisfies $\sum_{j \in {\cal{J}}_i }m_{ij}^{*}=1$, $\forall i  \in {\cal{N}}$.
\end{theorem}

Theorem \ref{theoremopcmp} can be proved by using the same method in \cite[Lemma~2]{Chin2015Power}.
Thus, the proof of Theorem \ref{theoremopcmp} is omitted.
For the convenience of analysis,  we substitute $\pmb{r} = \pmb{D}$ into Problem (\ref{max1_1}) in the following of this section.
\subsection{Distributed Time Allocation and Power Control}
We introduce $q_i$ as the average transmission power of BS $i$, which is given by
\begin{equation}\label{eqq}
q_i=\sum_{j \in {\cal{J}}_i }m_{ij} p_{ij}.
\end{equation}
Applying this result to Equation (\ref{eq2}) yields
\begin{eqnarray}\label{eq4}
m_{ij}
=
\frac{D_{ij}}
{B\log_2\left(1+ \frac{p_{ij} g_{ij}} {\sum_{k \in {\cal{N}}\setminus \{i\}}  q_k g_{kj} +\sigma ^2} \right)}.
\end{eqnarray}
Based on (\ref{eq4}), we use $m_{ij}$ to represent $p_{ij}$,
\begin{eqnarray}\label{eq4_2}
p_{ij}&&\!\!\!\!\!\!\!\!\!\!
=\frac{\sum_{k \in {\cal{N}}\setminus \{i\}}  q_k g_{kj} +\sigma ^2}
{g_{ij}}
\left( \text e^{ \frac{\ln(2)D_{ij}} {B m_{ij}}  }-1
\right)
\nonumber\\
 &&\!\!\!\!\!\!\!\!\!\!
 \triangleq h_{ij}(m_{ij},\pmb{q}_{-i}, D_{ij}), \quad\forall i  \in {\cal{N}}, \forall j \in {\cal{J}}_i,
\end{eqnarray}
where $\pmb{q}_{-i}=(q_1, \cdots, q_{i-1}, q_{i+1}, \cdots, q_N)^T$.
Plugging (\ref{eq4_2}) into (\ref{eqq}), we have
\begin{eqnarray}\label{eq5}
q_{i}&&\!\!\!\!\!\!\!\!\!\!
=\sum_{j \in {\cal{J}}_i }
m_{ij} h_{ij}(m_{ij},\pmb{q}_{-i}, D_{ij}), \quad\forall i  \in {\cal{N}}.
\end{eqnarray}

By adopting the new power notation $\pmb q= (q_1, \cdots, q_N)^T$, we have the following theorem.

\begin{theorem}
If the feasible set of Problem (\ref{max1_1}) is not empty, sum power minimization Problem (\ref{max1_1}) is equivalent to the following problem,
\begin{subequations}\label{max1_2}
\begin{align}
\mathop{\min}_
{\substack{
\pmb{0} \leq \pmb{m}\\
\pmb{0} \leq \pmb{q} \leq \pmb{Q}^{\max}
}
}
\:
&\sum_{i \in {\cal{N}}} q_i
\\
\textrm{s.t.}\quad \:\:\:
& \sum_{j \in {\cal{J}}_i } m_{ij} h_{ij}(m_{ij},\pmb{q}_{-i}, D_{ij})\leq q_i,\quad \forall i  \in {\cal{N}}
\\
&\sum_{j \in {\cal{J}}_i } m_{ij} \leq 1, \forall i  \in {\cal{N}},
\end{align}
\end{subequations}
where $\pmb{Q}^{\max}=(P_1^{\max}, \cdots, P_N^{\max})^T$.
\end{theorem}

\itshape \textbf{Proof:}  \upshape Please refer to Appendix C. \hfill $\Box$

According to Theorem 3, the optimal solution of Problem (\ref{max1_1}) can be obtained by solving Problem (\ref{max1_2}).
However, the globally optimal solution of Problem (\ref{max1_2}) is also difficult to be effectively obtained due to nonconvex constraints (\ref{max1_2}b).
To solve Problem (\ref{max1_2}), we provide a novel distributed algorithm.
Denote $\pmb{m}_i=(m_{i (J_{i-1}+1)}, m_{i (J_{i-1}+2)}, \cdots, m_{i J_{i}})^T$, which can be viewed as the load vector of BS $i$.
Let $\pmb m_{-i}=(\pmb m_1^T, \cdots, \pmb m_{i-1}^T, \pmb m_{i+1}^T, \cdots, \pmb m_N^T)^T$.
With $\pmb q_{-i}$ and $\pmb m_{-i}$ fixed, BS $i$ should solve the following optimization problem,
\begin{subequations}\label{max1_2_2}
\begin{align}
\!\!\!\mathop{\min}_
{\substack{
\pmb{0} \leq \pmb{m}_i\\
{0} \leq  {q}_i \leq {P}^{\max}_i
}
}
\:
& q_i
\\
\textrm{s.t.}\quad\:\:
&\!\! \sum_{j \in {\cal{J}}_i }m_{ij} h_{ij}(m_{ij},\pmb{q}_{-i}, D_{ij})\leq q_i
\\
&\!\!\!  \sum_{l \in {\cal{J}}_k } \!\!m_{kl} h_{kl}(m_{kl},\pmb{q}_{-k}, D_{kl})\!\leq\! q_k, \forall k\!\in\! {\cal{N}}\!\setminus \!\{i\}\!\!\!\!\!
\\
&\!\!\! \sum_{j \in {\cal{J}}_i } m_{ij} \leq 1.
\end{align}
\end{subequations}

Substituting (\ref{eq4_2}) into constraints (\ref{max1_2_2}c), we have
\begin{equation*}\label{eq4_3}
\sum_{l \in {\cal{J}}_k } m_{kl}\frac{q_i g_{il}\!+\!\sum_{n \in {\cal{N}}\setminus \{k,i\}}  q_n g_{nl} +\sigma ^2}
{g_{kl}}
\left(\!\text  e^{\frac {\ln(2)D_{kl}}{B m_{kl}} }\!-\!1
\!\right)
\leq q_k,
\end{equation*}
which is a linear constraint with $q_i$, $\forall k \in \mathcal N\setminus \{i\}$ .
Combining (\ref{max1_2_2}b) and (\ref{max1_2_2}c), we find that for any optimal solution to Problem (\ref{max1_2_2}),
(\ref{max1_2_2}b) holds with equality, as otherwise (\ref{max1_2_2}a) can be improved, contradicting that the solution is optimal.
Thus, the optimal $q_i^*$ of Problem (\ref{max1_2_2}) satisfies
\begin{equation}\label{eq4_4}
q_i^* =\mathop{\min}\limits_{\substack{\pmb 0 \leq\pmb{m}_i\\
\sum_{j \in \mathcal J_i}m_{ij}\leq 1}}
\quad
\sum_{j \in \mathcal J_i}m_{ij} h_{ij}(m_{ij},\pmb{q}_{-i}, D_{ij})
\end{equation}

From (\ref{eq4_4}), we can observe that the optimal power $q_i$ of BS $i$ is only determined by  power
$\pmb q_{-i}$ of all other BSs and the load vector $\pmb{m}_i$ of BS $i$.
In our distributed power optimization design, the load of each user is determined by its served BS.
Thus, given other BSs' power $\pmb q_{-i}$, BS $i$ should solve the following optimization problem to obtain the optimal $q_i$.
\begin{subequations}\label{max1_3}
\begin{align}
\mathop{\min}_{\pmb{0}\leq\pmb{m}_i}\quad
&\sum_{j \in {\cal{J}}_i }
a_{ij} m_{ij}
\left( \text e^\frac{b_{ij}}{ m_{ij}} -1 \right)
\\
\textrm{s.t.}\qquad \!\!\!
&\sum_{j \in {\cal{J}}_i }
m_{ij}\leq 1,
\end{align}
\end{subequations}
where $a_{ij}=\frac{\sum_{k \in {\cal{N}}\setminus \{i\}}  q_k g_{kj} +\sigma ^2}
{g_{ij}}$, $b_{ij}=\frac {\ln(2)D_{ij}}  {B} $, $\forall j \in {\cal{J}}_i$.
Noth that variable $q_i$ cannot be directly found in power minimization Problem (\ref{max1_3}).
To the optimal solution of problem (\ref{max1_3}), $q_i$ equals to the objective function (\ref{max1_3}a).

According to (\ref{appendixAeq1}) in Appendix A,
it can be verified that Problem (\ref{max1_3}) is a convex problem.
The Lagrangian function of Problem (\ref{max1_3}) is
\begin{equation*}\label{eq9}
\mathcal L_1(\pmb{m}_i, \lambda_i)=
\sum_{j \in {\cal{J}}_i }
a_{ij} m_{ij}
\left(\text  e^{\frac{b_{ij}} {m_{ij}} }-1
\right) + \lambda_i \left(\sum_{j \in {\cal{J}}_i }
m_{ij}-1\right),
\end{equation*}
where $\lambda_i$ is the non-negative Lagrange multiplier associated with constraint (\ref{max1_3}b).

Since Problem (\ref{max1_3}) is similar to Problem (\ref{minimalTransmitPowerOpt}) in Appendix A, the optimal solution of Problem (\ref{max1_3}) can be obtained by using the same method in Appendix A.
Here, we directly give the results and the details can be found in Appendix A.
Define function $u(x)= x e^x -e^x +1$, and $u^{-1}(x)$ is the inverse function of $u(x)$, $x \geq 0$.
The Lagrangian variable $\lambda_i$ satisfies the following equation,
\begin{equation}\label{eq11}
1 =\sum_{j \in {\cal{J}}_i }
\frac{b_{ij}} { u^{-1} \left(\frac { \lambda_i} {a_{ij}} \right)} \triangleq  \hat u_i(\lambda_i) ,
\end{equation}
and the unique value of $\lambda_i$ can be obtained by using the bisection method.
Having obtained the Lagrangian variable $\lambda_i$, the optimal $\pmb m_i$ is calculated as
\begin{equation}
m_{ij}=\frac{b_{ij}} { u^{-1} \left(\frac { \lambda_i} {a_{ij}} \right)}, \quad\forall j \in \mathcal J_i.
\end{equation}

In the following, we provide a distributed time allocation and power control  algorithm to solve sum power optimization Problem (\ref{max1_1}) in Algorithm 1.
\begin{algorithm}[h]
\caption{Distributed Time Allocation and Power Control for Power Minimization (DTAPC-PM)}
\label{alg:Framwork1}
\begin{algorithmic}[1]
\State Initialize $q_i^{(0)}=P_i^{\max}$, $\forall i \in \mathcal N$.
Set iteration number $n=1$, and maximal iteration number $N_{\max}$.
\For{$i=1, 2, \cdots, N$}
\State With power $\pmb q_{-i}^{(n-1)}$ fixed, calculate $a_{ij}^{(n-1)}=\frac{\sum_{k \in {\cal{N}}\setminus \{i\}}  q_k^{(n-1)} g_{kj} +\sigma ^2} {g_{ij}}$, $\forall j \in {\cal{J}}_i$;
\State Use the bisection method to obtain the optimal $\lambda_i^{(n)}$ such that $\hat u_i(\lambda_i ^{(n)})=1$;
\State Update $m_{ij}^{(n)}={b_{ij}} \left/{ u^{-1} \left(\frac { \lambda_i^{(n)}} {a_{ij}^{(n-1)}} \right)}\right.$, $\forall j \in {\cal{J}}_i$,
$q_{i}^{(n)} =\sum_{j \in {\cal{J}}_i }
m_{ij}^{(n)} h_{ij}(m_{ij}^{(n)},\pmb{q}^{(n-1)}_{-i}, D_{ij})$;
\EndFor
\State  
If $n > N_{\max}$ or objective function (\ref{max1_2}a) converges,
output $\pmb m^*=\pmb m^{(n)}$, $p_{ij}^*=h_{ij}(m_{ij}^{(n)},\pmb{q}_{-i}^{(n)}, D_{ij}), \forall i  \in {\cal{N}}, \forall j \in {\cal{J}}_i$, $\pmb r^*=\pmb D$, and
terminate.
Otherwise, set $n=n+1$ and go to step 2.
\end{algorithmic}
\end{algorithm}

\subsection{Convergence and Global Optimality of DTAPC-PM}
To show the convergence and global optimality of DTAPC-PM algorithm,
we recap the standard interference function introduced in \cite{yates1995framework}.
Consider an arbitrary interference function $\pmb{I}(\pmb{q})=(I_1(\pmb{q}), \cdots, I_N(\pmb{q}))$, we say $\pmb{I}(\pmb{q})$ is a standard interference function if for all $\pmb{q} \geq \pmb{0}$, the following properties are satisfied.

1) Positivity: $\pmb{I}(\pmb{q}) > \pmb{0}$.

2) Monotonicity: If $\pmb{q}^{(1)} \geq \pmb{q}^{(2)}$, then $\pmb{I}(\pmb{q}^{(1)}) \geq \pmb{I}(\pmb{q}^{(2)})$.

3) Scalability: For all $\alpha >1$, $\alpha \pmb{I}(\pmb{q}) >\pmb{I}(\alpha \pmb{q})$.

Denote the solution of Problem (\ref{max1_3}) as $v_i(\pmb{q})$, $\forall i \in \cal{N}$.
Letting $\pmb{v}=(v_1, \cdots, v_N)$, we have the following theorem.
\begin{theorem}\label{theorempower1}
$\pmb{v}(\pmb{q})$ is a standard interference function.
\end{theorem}

\itshape \textbf{Proof:}  \upshape Please refer to Appendix D. \hfill $\Box$

Based on Theorem 4, we have the following corollaries.

\begin{coro}
If there exists $\pmb q$ such that $\pmb q \geq \pmb v(\pmb q)$, the iterative fixed-point method $\pmb q^{(i+1)}=\pmb v(\pmb q^{(i)})$
will converge to the unique fixed point $\pmb q^* =\pmb v(\pmb q^*)$ with any initial point $\pmb q^{(0)}$.
\end{coro}

\itshape \textbf{Proof:}  \upshape Please refer to \cite[Theorem~2]{yates1995framework}.
\hfill $\Box$
\begin{coro}
Problem (\ref{max1_2}) is feasible, if and only if there exists $\pmb q^*=\pmb v(\pmb q^*)$ and $\pmb q^* \leq \pmb Q^{\max}$, where $\pmb Q^{\max}=(P_1^{\max}, \cdots, P_N^{\max})$.
\end{coro}

\itshape \textbf{Proof:}  \upshape  Please refer to Appendix E.
\hfill $\Box$
\begin{coro}
If Problem (\ref{max1_2}) is feasible, the optimal solution ($\pmb m^*, \pmb q^*$) of Problem (\ref{max1_2}) is unique and $\pmb q^*=\pmb v(\pmb q^*)$.
\end{coro}

\itshape \textbf{Proof:}  \upshape Please refer to Appendix F.
\hfill $\Box$
\begin{coro}
If Problem (\ref{max1_2}) is feasible, the optimal $\pmb q^*$ of Problem (\ref{max1_2}) is component-wise minimum in the sense that any other feasible solution ($\pmb m', \pmb q'$) that satisfies (\ref{max1_2}b) and (\ref{max1_2}c) must satisfy $\pmb q' \geq \pmb q^*$.
\end{coro}

Since Corollary 4 can be easily proved from Corollary 1 and Corollary 2, the proof of Corollary 4 is omitted.

\textit{Remark}: In Algorithm 1, each BS solves a convex problem in each iteration.
According to \cite{yates1995framework}, when the iterative relation of the algorithm is a standard interference function and the problem is feasible, the algorithm always converges and the convergent solution is globally optimal.
Consequently, Algorithm 1 converges to the globally optimal solution based on the property of a standard interference function.
As a standard interference function, the iterative relation of Problem (\ref{max1_3}) satisfies three properties.
The positivity property should be satisfied since the transmission power of each BS should be positive to meet the positive rate demands of all users.
When the transmission power of some BSs increases, the transmission power of all BSs should be accordingly increased to meet the rate demands for all users, which implies the monotonicity property.
The scalability property states that the scaled input transmission power of all BSs results in upper-bounded output transmission power of all BSs.

\subsection{Complexity Analysis}
\label{companapm}
For simplicity of analysis, it is assumed that the number of users in each cell is $K$.
For the DTAPC-PM algorithm, in each iteration the major complexity lies in the computation of Lagrange multiplier $\lambda_i$.
The computation of $\lambda_i$ involves the inverse function $u^{-1}(x)$ in (\ref{eq11}),
and the complexity of computing $u^{-1}(x)$
is $\mathcal O(\log_2(1/\epsilon_0))$ by using bisection method for accuracy $\epsilon_0$.
To solve $\lambda_i$ from (\ref{eq11}), the complexity is $\mathcal O(K  \log_2(1/\epsilon_0)\log_2(1/\epsilon_1))$ with $\epsilon_1$ as the accuracy required for bisection method.
Thus, the total complexity of the DTAPC-PM is $\mathcal O(L_{\text{PM}}NK^2 \log_2(1/\epsilon_0)\log_2(1/\epsilon_1)) )$, where $L_{\text{PM}}$ denotes the total number of iterations of the DTAPC-PM algorithm.
For the optimal power vector for power minimization (OPV-PM) algorithm in \cite{Chin2015Power}, the main computational complexity lies in the bisection search of power, which involves a complexity of $\mathcal O(K \log_2(1/\epsilon_2))$ for accuracy $\epsilon_2$.
Then, the total complexity of the OPV-PM is $\mathcal O(L_{\text{OPV}}NK  \log_2(1/\epsilon_2))$, where $L_{\text{OPV}}$ denotes the total number of iterations by using the OPV-PM algorithm.
According to Fig.~9 in Section VI.B, the values of $L_{\text{PM}}$ and $L_{\text{OPV}}$ are small. The typical numbers of iterations are $L_{\text{PM}}$=$L_{\text{PM}}$=10 from Fig.~9.

\subsection{Implementation Method}
To successfully implement the DTAPC-PM algorithm, each BS $i$ needs to compute load vector $\pmb m_i$, which includes the coefficients $a_{ij}$, $\forall j \in \mathcal J_i$.
Since $a_{ij}=\frac{\sum_{k \in {\cal{N}}\setminus \{i\}}  q_k g_{kj} +\sigma ^2}
{g_{ij}}$,
the denominator $g_{ij}$ can be estimated by BS $i$ according to channel reciprocity.
For the numerator of $a_{ij}$, ${\sum_{k \in {\cal{N}}\setminus \{i\}}  q_k g_{kj} +\sigma ^2}$ is the total interference and noise power of user $j$ served by BS $i$ and the numerator of $a_{ij}$ can be calculated by user $j$.
Thus, the numerator of $a_{ij}$ can be obtained by the message sent from user $j$.
It is assumed that each BS knows the rate demands of its served users and the channel gains between the BS and its served users remain unchanged during the resource scheduling time.
In summary, every user calculates the value of received total interference power and sends this message to its served BS, then each BS $i$ estimates the channel gains between its served users and BS $i$.
Moreover, each BS $i$ calculates the optimal $\pmb m_i$ of Problem (\ref{max1_3}), and $p_{ij}=a_{ij} ( \text e^{b_{ij}/ m_{ij} }-1 ), \forall j \in \mathcal J_i$.
Each BS updates its load vector and transmission power vector until the total interference power of each user converges.

\subsection{Comparison with the Distributed Algorithm Using Alternating Directions Method of Multipliers}

To solve the sum power optimization Problem (3) with $V(\pmb m, \pmb p, \pmb r)=\sum_{i \in \mathcal N} \sum_{j\in\mathcal J_i}$ $m_{ij}p_{ij}$, there are two advantages in Algorithm 1:
\begin{enumerate}
  \item The globally optimal solution of Problem (3) with $V(\pmb m, \pmb p, \pmb r)=\sum_{i \in \mathcal N} \sum_{j\in\mathcal J_i}m_{ij}p_{ij}$ is obtained via Algorithm 1.
  \item Algorithm 1 can be implemented in practice since each BS can calculate the its own strategy according to the channel gains between this BS and its served users and the messages of total interference and noise power sent from its served users.
\end{enumerate}
For the distributed algorithm using alternating directions method of multipliers (ADMM) \cite{6484993,6156468,6748974}, it is hard to obtain the globally optimal solution of Problem (3) with $V(\pmb m, \pmb p, \pmb r)=\sum_{i \in \mathcal N} \sum_{j\in\mathcal J_i}m_{ij}p_{ij}$ due to that Problem (3) is nonconvex.
Besides, since the power vectors of different BSs in the Lagrangian of Problem (3) cannot be decoupled, it is difficult to establish the distributed algorithm using ADMM.

\section{Sum Rate Maximization Problem for Multi-cell Networks}

In this section, based on the results about sum rate maximization for single-cell networks, we solve the sum rate optimization Problem (\ref{max1_1}) with $V(\pmb m, \pmb p, \pmb r)=-\sum_{i \in \mathcal N} \sum_{j\in\mathcal J_i} r_{ij}$ for multi-cell networks.
We establish a distributed time allocation and power control algorithm, and provide the proof of convergence for this distributed algorithm.

\subsection{Distributed Time Allocation and Power Control}
\label{section_rate_1}
Due to nonconvex constraints (\ref{max1_1}b) and (\ref{max1_1}c),
the globally optimal solution of Problem (\ref{max1_1}) is difficult to be effectively obtained.
To solve Problem (\ref{max1_1}), we provide a novel distributed algorithm to obtain a suboptimal solution.
Denote $\pmb{p}_i=(p_{i (J_{i-1}+1)}, p_{i (J_{i-1}+2)}, \cdots, p_{i J_{i}})^T$,
$\pmb{r}_i=$ $(r_{i (J_{i-1}+1)}, r_{i (J_{i-1}+2)}, \cdots, r_{i J_{i}})^T$, which can be viewed as the power and rate vector of BS $i$, respectively.
Let $\pmb p_{-i}=(\pmb p_1^T, \cdots, \pmb p_{i-1}^T, \pmb p_{i+1}^T, \cdots, \pmb p_N^T)^T$, $\pmb r_{-i}=(\pmb r_1^T, \cdots, \pmb r_{i-1}^T, \pmb r_{i+1}^T, \cdots, \pmb r_N^T)^T$.
Plugging (\ref{eq2}) into (\ref{max1_1}),
BS $i$ should solve the following optimization problem with given $\pmb m_{-i}$, $\pmb p_{-i}$ and $\pmb r_{-i}$,
\begin{subequations}\label{ratemax1_2_2}
\begin{align}
\mathop{\min}_
{\substack{
\pmb{0} \leq \pmb{m}_i,
\pmb {0} \leq  \pmb {p}_i
}
}
\:
& -\sum_{j\in\mathcal J_i} B m_{ij}\log_2\left(1+  {p_{ij} \bar g_{ij}}   \right)
\\
\textrm{s.t.}\quad\:\:
&B m_{ij}\log_2\left(1\!+ \! {p_{ij} \bar g_{ij}} \right) \!\geq\! D_j, \quad \forall j\!\in\! \mathcal J_i\\
&B m_{kl}\log_2\!\left(\!\!1\!+ \!\frac{p_{kl} g_{kl}} {E_{kli}+\sum_{j \in {\cal{J}}_i }m_{ij} p_{ij} g_{il}}\! \!\right) \!\geq\! r_{kl},
\nonumber\\
&\qquad\qquad\qquad \qquad
\quad \forall k\!\in\!\mathcal N\setminus \{i\}, l\!\in\! \mathcal J_k\\
&\sum_{j \in {\cal{J}}_i }m_{ij} p_{ij} \leq P_i^{\max},  \\
& \sum_{j \in {\cal{J}}_i } m_{ij} \leq 1
\end{align}
\end{subequations}
where $\bar g_{ij\!}=\!{g_{ij}} \left/\left(\!{\sum_{k \in {\cal{N}}\setminus \{i\}} \sum_{l \in {\cal{J}}_k} m_{kl} p_{kl} g_{kj} \!+\!\sigma ^2}\!\right)\right.$, $E_{kli}=$ $\sum_{s \in {\cal{N}}\setminus \{i,k\}} \sum_{t \in {\cal{J}}_s} m_{st} p_{st} g_{sl} +\sigma ^2$, $\forall k\in\mathcal N\setminus \{i\}, l\in \mathcal J_k$.
To solve nonconvex Problem (\ref{ratemax1_2_2}), we introduce a set of new non-negative variables: $\bar p_{ij}=m_{ij}p_{ij}$, $\forall j \in \mathcal J_i$.
Then, Problem (\ref{ratemax1_2_2}) is equivalent to the following convex problem.
\begin{subequations}\label{ratemax1_2_3}
\begin{align}
\mathop{\min}_
{\substack{
\pmb{0} \leq \pmb{m}_i,
\pmb {0} \leq  \pmb {\bar p}_i
}
}
\:
& -\sum_{j\in\mathcal J_i}B m_{ij}\log_2\left(1+ \frac{\bar p_{ij} \bar g_{ij}} {m_{ij}} \right)
\\
\textrm{s.t.}\quad\:\:
&B m_{ij}\log_2\left(1\!+ \!\frac{\bar p_{ij} \bar g_{ij}} {m_{ij}} \right) \!\geq\! D_j, \quad \forall j\!\in\! \mathcal J_i\\
&\sum_{j \in {\cal{J}}_i }\bar p_{ij} \leq \bar P_i^{\max},  \\
& \sum_{j \in {\cal{J}}_i } m_{ij} \leq 1
\end{align}
\end{subequations}
where $\pmb{\bar p}_i=(\bar p_{i (J_{i-1}+1)}, \bar p_{i (J_{i-1}+2)}, \cdots, \bar p_{i J_{i}})^T$, $\bar P_i^{\max}=\min\{P_i^{\max},\min_{k \in \mathcal N\setminus \{i\}, l \in  \mathcal J_k} \{\bar P_{kli}^{\max}\}\}$, and
\begin{equation}\label{ratemax1_2_4}
\bar P_{kli}^{\max}=\frac{\bar p_{kl}g_{kl}}
{m_{kl} g_{il}\left(2^{\frac{r_{kl}}{B m_{kl}}}-1\right)}-\frac{E_{kli}}{g_{il}}, \quad  k \in \mathcal N\setminus \{i\}, l \in  \mathcal J_k.
\end{equation}
It is interesting to observe that Problem (\ref{ratemax1_2_3}) has the same structure of sum rate optimization Problem (\ref{OptimizationSingleCell}) in single-cell networks.
Thus, Problem (\ref{ratemax1_2_3}) can be effectively solved by using the same method in Appendix B.
Define $\bar g_{ij}$ as the effective channel gain of user $j$, which is the ratio of channel gain and inter-cell interference pulsed noise power.
From Appendix B, we can obtain the following theorem.
\begin{theorem} \label{theoremmulticellrate}
To maximize the sum rate of multi-cell networks, it is optimal for each BS to transmit with full time resources, i.e., $\sum_{j\in\mathcal J_i} m_{ij}^*=1$, $\forall i \in \mathcal N$.
Besides, the optimal power allocation strategy for each BS is to allocate
the additional power to the user with the best effective channel gain,
while other users are allocated with the minimal power to
maintain their minimal rate requirements.
\end{theorem}

In the following, we provide a distributed time allocation and power control  algorithm to solve sum rate optimization Problem (\ref{max1_1}) in Algorithm 3.
\begin{algorithm}[h]
\caption{Distributed Time Allocation and Power Control for Rate Maximization (DTAPC-RM)}
\label{alg:Framwork1}
\begin{algorithmic}[1]
\State Initialize feasible solution ($\pmb{m}^{(0)}$, $\pmb{p}^{(0)}$, $\pmb{r}^{(0)}$).
Set iteration number $n=1$, and maximal iteration number $N_{\max}$.
\For{$i=1, 2, \cdots, N$}
\State With $\pmb{m}_{\!-\!i}^{(n\!-\!1)}\!\!=\!(\pmb{m}_1^{(n)}; \!\cdots\!, \pmb{m}_{i\!-\!1}^{(n)};
\pmb{m}_{i\!+\!1}^{(n\!-\!1)}; \!\cdots\!; \pmb{m}_N^{(n\!-\!1)})$,
 $\pmb{p}_{-i}^{(n\!-\!1)}\!=\! (\pmb{p}_1^{(n)};  \!\cdots\!; \pmb{p}_{i\!-\!1}^{(n)};
\pmb{p}_{i\!+\!1}^{(n\!-\!1)}; \!\cdots\!; \pmb{p}_N^{(n\!-\!1)})$, and
 $\pmb{r}_{-i}^{(n-1)}=$ $(\pmb{r}_1^{(n)};  \cdots; \pmb{r}_{i-1}^{(n)};
\pmb{r}_{i+1}^{(n-1)}; \cdots; \pmb{r}_N^{(n-1)})$
fixed, obtain $\pmb m_i^{(n)}$, $\pmb {\bar p}_i^{(n)}$  by solving convex Problem (\ref{ratemax1_2_3});
\State $p_{ij}^{(n)}=\frac{\bar p_{ij}^{(n)}}{m_{ij}^{(n)}}$, $r_{ij}^{(n)}=B m_{ij}^{(n)}\log_2\left(1+ \frac{\bar p_{ij}^{(n)} \bar g_{ij}} {m_{ij}^{(n)}} \right)$, $\forall j \in \mathcal J_i$.
\EndFor
\State
If $n > N_{\max}$ or objective function (\ref{max1_1}a) converges, output $\pmb m^*=(\pmb m_1^{(n)}; \cdots; \pmb m_N^{(n)})$, $\pmb p^*=(\pmb p_1^{(n)}; \cdots; \pmb p_N^{(n)})$, $\pmb r^*=(\pmb r_1^{(n)}; \cdots; \pmb r_N^{(n)})$, and
terminate.
Otherwise, set $n=n+1$ and go to step 2.
\end{algorithmic}
\end{algorithm}

\subsection{Convergence of DTAPC-RM}
\begin{theorem}\label{rate_convergence}
Assuming $N_{\max}\rightarrow \infty$, the sequence of load, power and rate vectors generated by the sequential updating DTAPC-RM algorithm will converge.
\end{theorem}
\itshape \textbf{Proof:}  \upshape
Please refer to Appendix G.
\hfill $\Box$
\subsection{Complexity Analysis}
Assume that the number of users in each cell is $K$.
For the DTAPC-RM algorithm, in each iteration the major complexity lies in solving convex Problem (\ref{ratemax1_2_3}).
To obtain the optimal solution of Problem (\ref{ratemax1_2_3}), the KKT conditions are solved in Appendix B.
To solve Lagrange variable $\gamma$, the complexity is $\mathcal O(K\log_2(1/\epsilon_3) \log(1/\epsilon_4))$, where $O(\log_2(1/\epsilon_3)$ is the complexity of solving (\ref{eqapp2_16}) with bisection method for accuracy $\epsilon_3$ and $O(\log_2(1/\epsilon_4)$ is the complexity of obtaining the inverse function $w^{-1}(x)$ in (\ref{eqapp2_11}) and (\ref{eqapp2_6}) with bisection method for accuracy $\epsilon_4$.
Thus, the complexity of solving convex Problem (\ref{ratemax1_2_3}) is $\mathcal O(K^2\log_2(1/\epsilon_3) \log(1/\epsilon_4))$.
As a result, the total complexity of the DTAPC-RM is $\mathcal O(L_{RM}NK^2\log_2(1/\epsilon_3) \log(1/\epsilon_4))$, where $L_{RM}$ denotes the total number of iterations of the DTAPC-RM algorithm.
From Fig.~10 in Section VI.B, the typical value of $L_{\text{RM}}$ is 10.

\subsection{Implementation Method}

To implement the proposed DTAPC-RM algorithm, each BS $i$ needs to update the load vector $\pmb m_i$ and power vector $\pmb {\bar p}_i$ by solving Problem (\ref{ratemax1_2_3}).
Solving Problem (\ref{ratemax1_2_3}) involves effective channel gain $\bar g_{ij}$ and maximal transmission power $\bar P_i^{\max}$.
Since  $\bar g_{ij}=\frac{g_{ij}}{{\sum_{k \in {\cal{N}}\setminus \{i\}}  q_k g_{kj} +\sigma ^2}}$,
numerator $g_{ij}$ is the channel gain between BS $i$ and its served user $j$.
Obviously, $g_{ij}$ can be detected by channel reciprocity.
Beside,
${\sum_{k \in {\cal{N}}\setminus \{i\}}  q_k g_{kj} +\sigma ^2}$ is the total interference power of user $j$ served by BS $i$, and the value of interference power can be detected by user $j$.
Since $\bar P_i^{\max}=\min\{P_i^{\max},\min_{k \in \mathcal N\setminus \{i\}, l \in  \mathcal J_k} \{\bar P_{kli}^{\max}\}\}$, $\bar P_{kli}^{\max}$ should be obtained before calculating $\bar P_i^{\max}$.
According to (\ref{ratemax1_2_4}), obtaining $\bar P_{kli}^{\max}$ involves $\bar p_{kl}$, $g_{kl}$, $r_{kl}$, $m_{kl}$, $g_{il}$ and $E_{kli}$.
Since $\bar p_{kl}$ and $m_{kl}$ respectively are the power spectral density and fraction of resources of BS $k$ allocated to its served user $l$, $\bar p_{kl}$ and $m_{kl}$ are known by BS $k$.
The channel gain $g_{kl}$ between BS $k$ and its served user $l$ can be estimated by channel reciprocity.
Achievable rate $r_{kl}$ of user $l$ served by BS $k$ can be known by BS $k$.
For cross channel gain $g_{il}$ between BS $i$ and user $l$ served by BS $k\neq i$, $g_{il}$ can be estimated at BS $i$ for receiving the pilot from user $l$ according to channel reciprocity.
To calculate $E_{kli}$,
\begin{eqnarray}
E_{kli}&&\!\!\!\!\!\!\!\!\!\!
=\sum_{s \in {\cal{N}}\setminus \{i,k\}} \sum_{t \in {\cal{J}}_s} \bar p_{st} g_{sl} +\sigma ^2
=I_{kl}-\sum_{j \in \mathcal J_i}\bar p_{ij} g_{il}, \nonumber
\end{eqnarray}
where $I_{kl}=\sum_{s \in {\cal{N}}\setminus \{k\}} \sum_{t \in {\cal{J}}_s} \bar p_{st} g_{sl} +\sigma ^2$ is the total interference and noise received from user $l$, and $\sum_{j \in \mathcal J_i}\bar p_{ij}$ is the sum power spectral density of BS $i$.
As a result, user $l\in \mathcal J_k$ sends its overall received interference and noise to BS $k$.
Then, having obtained the messages from users in $\mathcal J_k$, BS $k$ sends these obtained messages as well as its strategy ($\pmb{\bar p}_{k}$, $\pmb m_{k}$, $\pmb r_{k}$) to BS $i$.
As a result, BS $i$ calculates the optimal load vector $\pmb m_i$ and power vector $\pmb {\bar p}_i$ by solving Problem (\ref{ratemax1_2_4}).
Each BS updates its load vector and power vector until the total interference power of each user converges.

\section{Numerical Results}

\subsection{Optimization for A Single-Cell Network}
We consider a single-cell network with 9 users.
The path loss model is the NLOS scenario of Urban Micro cell and the standard deviation of shadow fading is $4$ dB \cite{access2010further}.
The noise power $\sigma^2=-174$ dBm/Hz, and the bandwidth of the system is $B=18$ MHz.
We assume equal rate demand for all users, i.e., $D_j=D$, $\forall j$.
We compare the following two algorithms: sum power minimization for a single cell (labeled as `PM-SC'), and sum rate maximization for a single cell (labeled as `RM-SC').

Fig. 2 to Fig. 5 illustrate the channel gains, time allocation, power control and rate distribution for different users, respectively.
In Fig.~3 to Fig.~5, we set $D=2.5$ Mbps and maximal power $P_{\max}=1$ W.
For simplicity, the power of each BS is measure in watt in the following.
Summing the fractions of time resources for all users in Fig.~3, we can find that the BS uses full time resources for both PM-SC and RM-SC.
Thus, to minimize sum power, or maximize sum rate, it is always optimal for the BS to use all the time resources, which verifies the theoretical findings in Theorem 1.
The channel for user 7 is better than user 8 according to Fig.~2, while the allocated fraction of time for user 7 is smaller than the allocated fraction of time for user 8 based on Fig.~3.
We can observe that the key to save energy or increase throughput is to allocate more time to the users with poor channel gains.
From Fig.~2 and Fig.~3, we can find that the majority of time resources are allocated to the user with highest channel gain for RM-SC, while most of time resources are allocated to the user with weakest channel gain for PM-SC.
This is due to the fact that a great number of time resources allocated to the user with highest channel gain can produce high data rate and these resources allocated to the user with weakest channel gain can significantly decrease the required power.
To maximize sum rate, we observe that the BS allocates highest power to the user with highest channel gain for RM-SC according to Fig.~4.
In Fig.~5, it is observed that all the users are set with minimal rate demands for PM-SC, while the user with highest channel gain is set with highest rate for  RM-SC.

\begin{figure}
\centering
\includegraphics[width=2.6in]{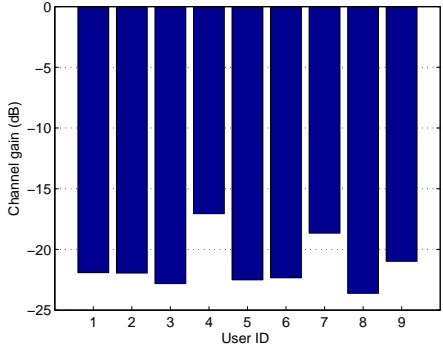}
\vspace{-1em}
\caption{Channel gains for different users.\label{fig6}}
\vspace{-1em}
\end{figure}

\begin{figure}
\centering
\includegraphics[width=2.6in]{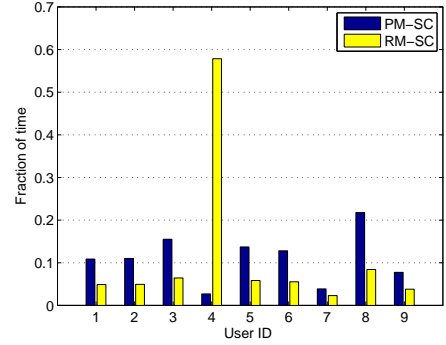}
\vspace{-1em}
\caption{Time allocation for different users.\label{fig3}}
\vspace{-1em}
\end{figure}

\begin{figure}
\centering
\includegraphics[width=2.6in]{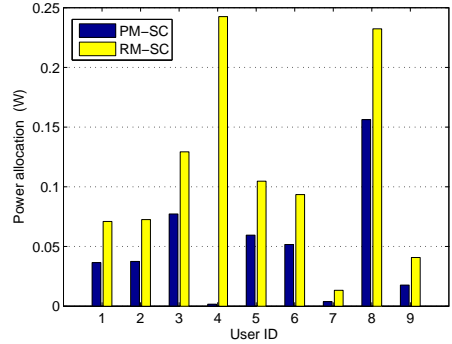}
\vspace{-1em}
\caption{Power control for different users.\label{fig4}}
\end{figure}

\begin{figure}
\centering
\includegraphics[width=2.6in]{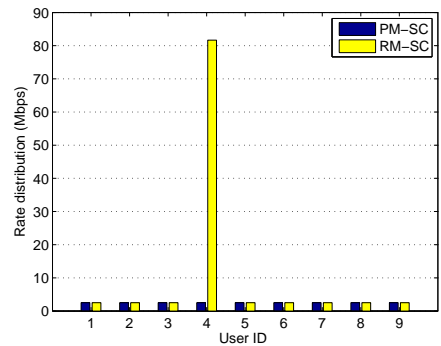}
\vspace{-1em}
\caption{Rate distribution for different users.\label{fig5}}
\vspace{-1em}
\end{figure}

\begin{figure}
\centering
\includegraphics[width=2.6in]{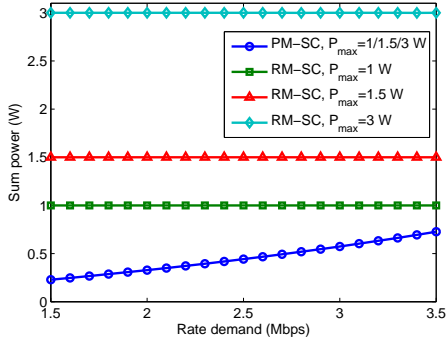}
\vspace{-1em}
\caption{Sum power performance under different rate demands.\label{fig7}}
\end{figure}

\begin{figure}
\centering
\includegraphics[width=2.6in]{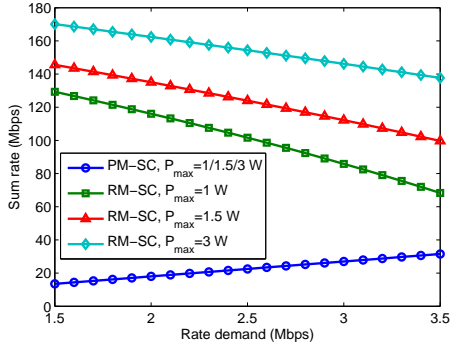}
\vspace{-1em}
\caption{Sum rate performance under different rate demands.\label{fig8}}
\vspace{-1em}
\end{figure}

The sum power and sum rate performance under different rate demands are shown in Fig.~\ref{fig7} and Fig.~\ref{fig8}, respectively.
From Fig.~\ref{fig7}, it is observed that sum power is always maximal for RM-SC, while sum power increases with rate demand for PM-SC.
This is because maximal power should be always used to maximize sum rate, and the increase of rate demand requires additional power for PM-SC.
Obviously, sum power of PM-SC outperforms RM-SC.
In Fig.~\ref{fig8}, sum rate of RM-SC is superior over PM-SC especially when the rate demand is low.
From Fig.~\ref{fig7} and Fig.~\ref{fig8}, we can conclude that RM-SC outperforms PM-SC in terms of sum rate at the increase of power consumption.

\subsection{Optimization for A Multi-Cell Network}
We consider a multi-cell network consisting of 15 cells and 450 users.
The three-sector antenna pattern is
used for each site
and the gain for the three-sector,
of which
3dB beamwidth in degrees
is 70 degrees, is 14dBi.
The path loss model is
the NLOS scenario of Urban Micro cell
and the standard deviation of shadow fading
is $4$ dB \cite{access2010further}.
Each user selects the BS with the highest channel gain (the user randomly chooses one BS if there would be multiple BSs with the highest channel gain).
We assume equal maximal transmission power (i.e., $P_i^{\max}=P^{\max}=10$ W, $\forall i \in \mathcal N$) for all BSs and equal rate demand (i.e., $D_{ij}=D$, $\forall i\in \mathcal N$, $j\in \mathcal J_i$) for all users.
We compare the performance of proposed DTAPC-PM and DTAPC-RM with the OPV-PM algorithm in \cite{Chin2015Power}, where uniform power is allocated to users in the same cell, and the WMMSE-RM algorithm to solve sum rate optimization (3) by iterative solving the time allocation problem and power control problem where the power control problem is solved via the minimization of weighted mean-square error (WMMSE) approach \cite{5756489}.

\begin{figure}
\centering
\includegraphics[width=2.6in]{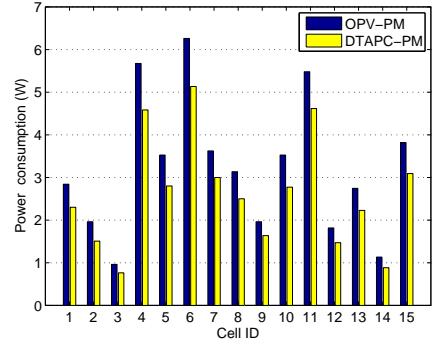}
\vspace{-1em}
\caption{Power consumption in each cell under $D=1$ Mbps. \label{fig13}}
\vspace{-1em}
\end{figure}

In Fig.~\ref{fig13}, we illustrate the power solutions of each BS by using OPV-PM and DTAPC-PM.
The proposed DTAPC-PM significantly outperforms the OPV-PM in terms of requiring lower power for every BS.
The total transmission power of all cells by the proposed DTAPC-PM is about 39\% lower than by OPV-PM.
This is due to the fact that OPV-PM assumes equal power allocation for users served by the same BS, while DTAPC-PM assumes unequal power allocation for users served by the same BS to further exploit multiuser diversity.

\begin{figure}
\centering
\includegraphics[width=2.6in]{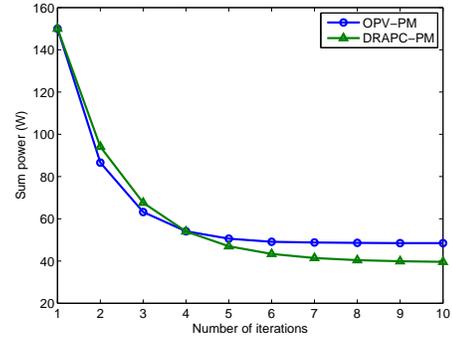}
\vspace{-1em}
\caption{Convergence behavior of OPV-PM and DTAPC-PM under $D=1$ Mbps.\label{fig14}}
\vspace{-1em}
\end{figure}

\begin{figure}
\centering
\includegraphics[width=2.6in]{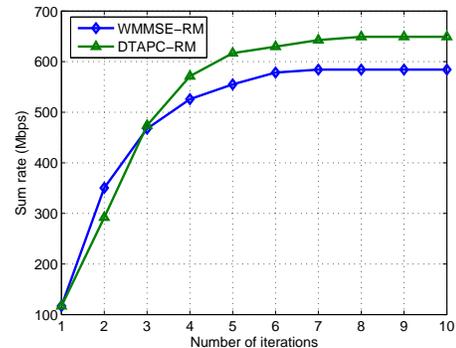}
\vspace{-1em}
\caption{Convergence behavior of MMSE-RM and DTAPC-RM under $D=1$ Mbps.\label{fig142}}
\vspace{-1em}
\end{figure}

Fig.~\ref{fig14} shows the convergence behavior of OPV-PM and DTAPC-PM.
It can be seen that both OPV-PM and DTAPC-PM monotonically increase and converge quickly.
Obviously, the sum power of DTAPC-PM outperforms OPV-PM.
This is due to that DTAPC-PM further exploits multiuser diversity by setting users served by the same BS with unequal power allocation, which requires more computations in each iteration than OPV-PM according to the complexity analysis in Section IV.D.
Thus, we can conclude that the DTAPC-PM achieves a significant performance gain at the cost of some additional computations.
Fig.~\ref{fig142} shows the convergence behavior of WMMSE-RM and DTAPC-RM.
It can be observed that the sum rate of DTAPC-RM is superior over WMMSE-RM.
For both WMMSE-RM and DTAPC-RM, the sum rate monotonically increases and converges rapidly, which makes the algorithms suitable for practical applications.

\begin{figure}
\centering
\includegraphics[width=2.6in]{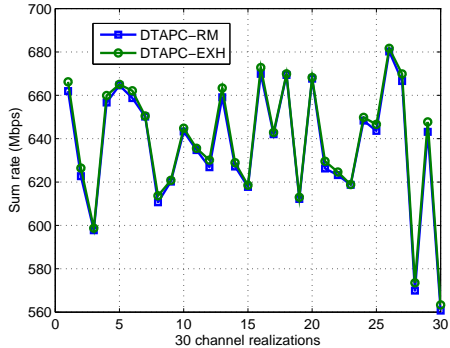}
\caption{Performance comparison of DTAPC-EXH and DTAPC-RM under $D=1$ Mbps.\label{fig17}}
\end{figure}

We try multiple starting points in the simulations to exhaustively obtain a near globally optimal solution.
We test 30 randomly generated channels shown in Fig.~\ref{fig17}, where DTAPC-EXH refers to the DTAPC-RM algorithm with 1000 starting points for each channel realization.
It can be seen that the sum rate of DTAPC-RM is almost the same as that of DTAPC-EXH, implying that the proposed DTAPC-RM approaches the near globally optimal solution.

\begin{figure}
\centering
\includegraphics[width=2.6in]{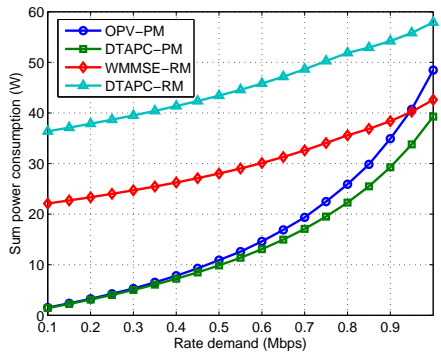}
\vspace{-1em}
\caption{Sum power performance under different rate demands for a multi-cell network.\label{fig15}}
\vspace{-1em}
\end{figure}

\begin{figure}
\centering
\includegraphics[width=2.6in]{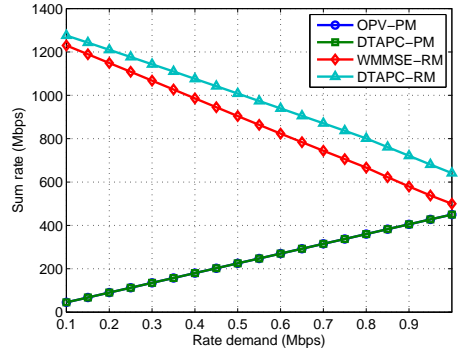}
\vspace{-1em}
\caption{Sum rate performance under different rate demands for a multi-cell network.\label{fig16}}
\end{figure}

The sum power, and sum rate performance under different rate demands for a multi-cell network are shown in Fig.~\ref{fig15} and Fig.~\ref{fig16}, respectively.
According to Fig.~\ref{fig15}, DTAPC-PM outperforms the other three algorithms and sum power is greatly reduced by using DTAPC-PM compared to OPV-PM when the rate demand is large.
It is also observed that sum power monotonically increases with the minimal rate demand for all algorithms.
Different from Fig.~\ref{fig7}, where sum power is always maximal to maximize sum rate in a single-cell network,
sum power monotonically increases with the minimal rate demand for DTAPC-RM in a multi-cell network.
This is because that maximal transmission power is not always optimal due to mutual interference among cells.
In Fig.~\ref{fig16}, sum rate of DTAPC-RM is largest among four algorithms.
The DTAPC-RM outperforms the WMMSE-RM especially when the rate demands are large.
Besides, sum rate monotonically decreases with the minimal rate demand for DTAPC-RM.
This is due to the fact that high rate demand requires large transmission power of each BS,
causing large mutual interference among cells and low achievable rate for users.

\section{Conclusions}
In this paper, we have optimized sum power and rate through time allocation and power control for multi-cell networks with load coupling.
We first obtain the globally optimal solutions for single-cell networks by solving KKT conditions.
Based on the results in single-cell networks, we propose two distributed time allocation and power control algorithms for optimization problems in multi-cell networks.
These two distributed algorithms can both be proved convergent, and interestingly the distributed algorithm for sum power minimization can be rigorously proved globally optimal.
Our analytical results suggest that maximal use of time resources is optimal to minimize sum power and maximize sum rate for both single-cell networks and multi-cell networks.
To minimize sum power for single-cell networks, we show that all users are allocated with minimal power to satisfy the rate demands.
To maximize sum rate for single-cell networks, we show that user with the best channel gain is allocated with additional power, while other users are allocated with minimal power to satisfy the rate demands.
These optimal conditions about power control strategy for single-cell networks are also applicable to multi-cell networks with channel gain replaced by effective channel gain, which is defined as the channel gain divided by the totally received interference power including inter-cell interference and noise power.
\appendices
\section{KKT Conditions of Sum Power Minimization Problem}
To minimize sum power, we find that constraint (\ref{OptimizationSingleCell}b) holds with equality for optimal solution, as otherwise (\ref{OptimizationSingleCell}a) can be improved by decreasing power.
Setting constraint (\ref{OptimizationSingleCell}b) with equality, we have
\begin{equation}\label{qfuntionOfm}
\bar p_{j}
=a_j m_j \left(\textrm{e}^{\frac {b_{j}} {m_{j}}}-1\right),\quad j =1, \cdots, M,
\end{equation}
where $a_j=\sigma^2/g_j$ and $b_j=(\ln 2) D_j /B$.
Substituting (\ref{qfuntionOfm}) into (\ref{OptimizationSingleCell}), we have the following equivalent problem,
\begin{subequations}\label{minimalTransmitPowerOpt}
\begin{align}
\mathop{\min}_{\pmb 0 \leq \pmb{\tilde m}} \quad\!\!
&\sum_{j=1}^M a_j m_j \left(\textrm{e}^{\frac {b_{j}}  {m_{j}}}-1\right)\\
\textrm{s.t.}\qquad \!\!\!\!\!\!\!
&\sum_{j=1}^{M}m_j \leq 1.
\end{align}
\end{subequations}
It can be proved that Problem (\ref{minimalTransmitPowerOpt}) is convex.
To show this,
\begin{equation}\label{appendixAeq1}
\frac {\text{d}^2 a_{j} m_{j}
( e^{b_{j}/ m_{j} }-1
) }
{ \text{d} m_{j}^2  }= \frac{b_{j}^2} {m_{j} ^3}
e^{\frac{b_{j}} {m_{j}}} >0, \quad \forall m_{j} >0,
\end{equation}
which indicates that the objective function (\ref{minimalTransmitPowerOpt}a) is convex.
The lagrangian of Problem (\ref{minimalTransmitPowerOpt}) is
\begin{equation}
\mathcal L_2(\pmb{\tilde m}, \lambda)=
\sum_{j \in {\cal{J}}_i }
a_{j}  m_j \left(\textrm{e}^{\frac {b_{j}}  {m_{j}}}-1\right) + \lambda \left(\sum_{j =1}^M
m_{ij}-1\right),
\end{equation}
where $\lambda$ is the non-negative Lagrange multiplier associated with constraint (\ref{minimalTransmitPowerOpt}b).
Since $\lim_{m_j \rightarrow 0+}  m_j \textrm{e}^{ {b_{j}} / {m_{j}}}=+\infty$, we observe that for the optimal solution of Problem (\ref{appendixAeq1}), $\pmb {\tilde m}> \pmb 0$.

According to \cite{boyd2004convex}, the KKT conditions of (\ref{minimalTransmitPowerOpt}) are
\begin{subequations}\label{KKT1}
\begin{align}
\frac{\partial \mathcal L_2}{\partial m_j}=
a_{j} \left(\textrm{e}^{\frac{b_{j}} {m_{j}}}-\frac{b_{j}} {m_{j}} \textrm e^{\frac{b_{j}} {m_{j}}} -1\right) + \lambda
&=0, \quad j =1, \cdots, M\\
\lambda \left(\sum_{j =1}^Mm_{j}-1\right)
&=0\\
\sum_{j=1}^{M}m_j-1
&\leq 0
\\
\lambda
\geq 0, \pmb {\tilde m} &> \pmb 0.
\end{align}
\end{subequations}

From (\ref{KKT1}a), we have
\begin{equation}\label{eqapp1}
\lambda=a_{j}\left (-\textrm{e}^{\frac{b_{j}} {m_{j}}}+\frac{b_{j}} {m_{j}} \textrm e^{\frac{b_{j}} {m_{j}}} +1\right).
\end{equation}
Define function
$u(x)= x \textrm e^x -\textrm e^x +1$, $x \geq 0$.
We have $u'(x)=x \textrm e^x >0$, $\forall x >0$.
Thus, function $u(x)$ is strictly increasing and $u(x) > u(0) =0$, $\forall x >0$.
Based on (\ref{eqapp1}), we have
\begin{equation}\label{eqapp1_3}
{m_{j}}=\frac {b_{j}}{u^{-1}\left({\frac{\lambda} {a_{j}}}\right)},\quad j =1, \cdots, M,
\end{equation}
where $u^{-1}(x)$ is the inverse function of $u(x)$.
According to (\ref{eqapp1}), $\lambda=a_j u\left({\frac{b_{j}} {m_{j}}}\right)>0$, which implies that (\ref{KKT1}c) holds with equality.
Plugging (\ref{eqapp1_3}) into (\ref{KKT1}c) yields
\begin{equation}\label{eqapp1_2}
1 =\sum_{j=1}^M
\frac{b_{j}} { u^{-1} (\frac { \lambda} {a_{j}} )}
\triangleq  \hat u(\lambda).\end{equation}

Equation (\ref{eqapp1_2}) has a unique solution $\lambda>0$.
Since $u(x)$ is strictly increasing,
inverse function $u^{-1}(x)$ is also strictly increasing in $(0, +\infty)$.
Thus, $\hat u(\lambda_i)$ is a strictly decreasing function in $(0, +\infty)$.
Owing to the fact that $\lim_{\lambda \rightarrow 0+} \hat u(\lambda) =+\infty$
and $\lim_{\lambda \rightarrow +\infty} \hat u(\lambda) =0$,
there exists one unique $\lambda$ satisfying $\hat u(\lambda)=1$, and the solution can be obtained by using the bisection method.
Having obtained the value of $\lambda$, the optimal $\pmb {\tilde m }$ and $\pmb {\bar p}$ can be obtained from (\ref{eqapp1_3}) and (\ref{qfuntionOfm}), respectively.

\section{KKT Conditions of Sum Rate Maximization Problem}
In the sum rate maximization problem, the Lagrangian function of Problem (\ref{OptimizationSingleCell}) is
\begin{eqnarray*}
\mathcal L_3(\pmb{\tilde m }, &&\!\!\!\!\!\!\!\!\!\!\pmb{\bar p}, \pmb{\alpha}, \beta ,\gamma, \pmb \mu, \pmb \nu)
=
-{\sum_{j=1}^M \frac{B m_j}{\ln2} \ln\left(1+\frac{g_j \bar p_j}{\sigma^2 m_j}\right)}
\\
 &&\!\!\!\!
+\sum_{j=1}^M \alpha_j \left({D_j-\frac{B m_j}{\ln2} \ln\left(1+\frac{g_j \bar p_j}{\sigma^2 m_j}\right)}\right)
\\
 &&\!\!\!\!
+\beta \left(\sum_{j=1}^{M}m_j -1\right)
+\gamma \left(\sum_{j=1}^M \bar p_j-P_{\max}\right)
\\
&&\!\!\!\!
-\sum_{j=1}^M \mu_j m_j -\sum_{j=1}^M \nu_j \bar p_j
,
\end{eqnarray*}
where $\pmb \alpha=(\alpha_1, \cdots, \alpha_M)^T$, $\pmb \alpha=(\alpha_1, \cdots, \alpha_M)^T$, $\pmb \alpha=(\alpha_1, \cdots, \alpha_M)^T$.
$\pmb \alpha$, $\beta$, $\gamma$, $\pmb \mu$, and $\pmb \nu$ are the non-negative Lagrange multipliers associated with corresponding constraints of Problem (\ref{OptimizationSingleCell}).
According to \cite{boyd2004convex}, the KKT conditions of (\ref{OptimizationSingleCell}) are
\begin{subequations}\label{KKT2}
\begin{align}
&\frac{\partial \mathcal L_3}
{\partial m_j}\!=\!
 \beta\!-\! {\frac{B(1\!+\!\alpha_j)}{\ln2}}  \!\ln\left(\!1\!\!+\!\!\frac{g_j \bar p_j}{\sigma^2 m_j}\!\right)
\!\!+\!\!
{\frac{B(1\!\!+\!\!\alpha_j){g_j \bar p_j}}{(\ln2)({\sigma^2 m_j\!\!+\!\!g_j\bar p_j})}}
\nonumber\\
&\qquad\qquad\qquad\qquad\qquad
-\mu_j= 0, \quad j =1, \cdots, M\\
&\frac{\partial \mathcal L_3}
{\partial \bar p_j}\!=\!-\frac{B{ g_j(1+\alpha_j)m_j}}{(\ln2){(\sigma^2 m_j\!+\!g_j\bar p_j)}} \!+\!\gamma
\!-\!\nu_j
\!=\!0, \quad j \!=\!1, \cdots, M\\
&\alpha_j \! \left(\!{D_j\!-\!\frac{B m_j}{\ln2} \ln\left(\!1\!+\!\frac{g_j \bar p_j}{\sigma^2 m_j}\!\right)}\!\right)
\!=\!0, \quad j \!=\!1, \cdots, M\\
&\beta \left(\sum_{j=1}^{M}m_j -1\right)
=0\\
&\gamma \left(\sum_{j=1}^M \bar p_j-P_{\max}\right)
=0\\
&\mu_jm_j=0, \nu_j\bar p_j=0,\quad j=1, \cdots, M \label{appendkktmunu}\\
&{D_j-\frac{B m_j}{\ln2} \ln\left(1+\frac{g_j \bar p_j}{\sigma^2 m_j}\right)}\leq 0,\quad j =1, \cdots, M\\
&\sum_{j=1}^{M}m_j-1
\leq 0\\
&{\sum_{j=1}^M \bar p_j}-P_{\max}
\leq 0\\
&\pmb \alpha \geq 0, \beta \geq 0,\gamma \geq 0, \pmb \mu \geq \pmb 0, \pmb \nu \geq \pmb 0,
\pmb {\tilde m } \geq \pmb 0, \pmb {\bar p}  \geq \pmb 0.
\end{align}
\end{subequations}

We first show that $\pmb \mu=\pmb \nu=\pmb 0$.
If there exists $\mu_j>0$, we can obtain $m_j=0$ from (\ref{appendkktmunu}).
Then, we have ${\frac{B m_j}{\ln2} \ln\left(1+\frac{g_j \bar p_j}{\sigma^2 m_j}\right)}=0<D_j$, which contradicts constraint
(\ref{OptimizationSingleCell}b).
Thus, we have $\pmb \mu=\pmb 0$ and $\pmb {\tilde m } > \pmb 0$.
Similarly, we can obtain $\pmb \nu=\pmb 0$ and $\pmb {\bar p} >\pmb 0$.

From (\ref{KKT2}a), we can obtain
\begin{equation}
\beta= {\frac{B(1+\alpha_j)}{\ln2}}  \left(\ln\left(1+\frac{g_j \bar p_j}{\sigma^2 m_j}\right)
-\frac{g_j \bar p_j}{\sigma^2 m_j+g_j \bar p_j} \right),
\end{equation}
for all $j=1,\cdots, M$.
To show that $\beta>0$, we introduce function
\begin{equation}
t(\bar p_j)=\ln\left(1+\frac{g_j \bar p_j}{\sigma^2 m_j}\right)
-\frac{g_j \bar p_j}{\sigma^2 m_j+g_j \bar p_j}, \quad \forall \bar p_j\geq 0.
\end{equation}
Then, we can obtain
\begin{equation}
t'(\bar p_j)=\frac{g_j^2\bar p_j}{\sigma^2 m_j+g_j\bar p_j} >0,
\end{equation}
which implies that $t(\bar p_j)$ is strictly increasing.
Thus,
\begin{equation}\label{eqapp_2_15}
\beta= {\frac{B(1+\alpha_j)}{\ln2}}t(\bar p_j)>{\frac{B(1+\alpha_j)}{\ln2}}t(0)=0.
\end{equation}

According to (\ref{KKT2}b), we have
\begin{equation}\label{eqapp2_1}
\gamma=\frac{B{ g_j(1+\alpha_j)m_j}}{(\ln2){(\sigma^2 m_j+g_j\bar p_j)}}>0,\quad j=1,\cdots, M.
\end{equation}
Rearrange the structure of (\ref{KKT2}b), we can obtain
\begin{equation}\label{eqapp2_2}
\frac{\sigma^2 m_j}{ {\sigma^2 m_j+g_j\bar p_j}}=\frac{(\ln2){\sigma^2\gamma}}{B{ g_j(1+\alpha_j)}}, \quad j=1,\cdots, M.
\end{equation}
Substituting the (\ref{eqapp2_2}) into (\ref{KKT2}a), we have
\begin{eqnarray}\label{eqapp2_3}
\beta&&\!\!\!\!\!\!\!\!\!\!=
{\frac{B(1+\alpha_j)}{\ln2}}  \left(\ln\frac{{B} g_j(1+\alpha_j)}{{(\ln2)}\sigma^2\gamma}
-1+\frac{{(\ln2)}\sigma^2\gamma}{{B} g_j(1+\alpha_j)}\right)
\nonumber\\
&&\!\!\!\!\!\!\!\!\!\!
\triangleq v (g_j, \alpha_j), \quad j=1,\cdots, M.
\end{eqnarray}
In the following, we show that function $v (g_j, \alpha_j)$ is strictly increasing with both $g_j$ and $\alpha_j$.
According to (\ref{eqapp2_3}), we have
\begin{equation}\label{appendixAeq3}
\frac{\partial v (g_j, \alpha_j)} {\partial g_j}= \frac{B(1+\alpha_j)}{(\ln2)g_j} - \frac{\sigma^2 \gamma} {g_j^2 }.
\end{equation}
From (\ref{eqapp2_2}),
\begin{equation}\label{appendixAeq4}
\frac{(\ln2){\sigma^2\gamma}}{B{ g_j(1+\alpha_j)}}=\frac{\sigma^2 m_j}{ {\sigma^2 m_j+g_j\bar p_j}}<1, \quad j=1,\cdots, M.
\end{equation}
Combining (\ref{appendixAeq3}) and (\ref{appendixAeq4}),
we have $\frac{\partial v (g_j, \alpha_j)} {\partial g_j}>0$.
Defining function $w(x)=x(\ln x -1 +1/x)$, (\ref{eqapp2_3}) can be reformulated as
\begin{equation}\label{appendixAeq2}
v (g_j, \alpha_j)=\frac{\sigma^2 \gamma}{g_j}w\left(
\frac{B{ g_j(1+\alpha_j)}}{(\ln2){\sigma^2\gamma}} \right), \quad j=1,\cdots, M.
\end{equation}
Since $w'(x)=\ln x >0, \forall x>1$,
we can obtain that $\frac{\partial v (g_j, \alpha_j)} {\partial \alpha_j}>0$.

In the practical systems,  the probability that $g_i = g_j$ for $i \neq j$ is always 0.
Thus, without loss of generality, it can be assumed that $g_j$ is arranged in a decreasing order, i.e.,
$g_1 > g_2 > \cdots > g_M$.
Due to the fact that $\beta=v(g_j, \alpha_j)$ and function $v(g_j, \alpha_j)$ is strictly increasing with both $g_j$ and $\alpha_j$, $j=1, 2, \cdots, M$, we have $\alpha_M>\cdots>\alpha_2>\alpha_1\geq0$.

For $\alpha_1$, we consider the two cases, $\alpha_1>0$ and $\alpha_1=0$.
If $\alpha_1>0$, every user transmits with minimal data rate according to (\ref{KKT2}c) and (\ref{KKT2}g).
As a result, the optimal value of sum rate maximization Problem (\ref{OptimizationSingleCell}) is $\sum_{j=1}^M D_j$.
Denote $\pmb {\bar D}=(D_1, \cdots, D_M)^T$ and the optimal value of sum power minimization Problem (\ref{minimalTransmitPowerOpt}) as $P_{\min}(\pmb {\bar D})$, which can be viewed as the minimal power to maintain the minimal rate demand vector $\pmb {\bar D}$.
Then, we can declare that $\alpha_1>0$ if and only if $P_{\min}(\pmb {\bar D})=P_{\max}$, as otherwise the objective function could be further improved with additional power.
When $\alpha_1>0$, the optimal $\pmb {\tilde m }$ and $\pmb {\bar p}$ can be obtained from (\ref{eqapp1_3}) and (\ref{qfuntionOfm}) as in Appendix A.

When $P_{\min}(\pmb {\bar D})<P_{\max}$, we can obtain $\alpha_1=0$.
Substituting $\alpha_1=0$ into (\ref{eqapp2_3}), we have
\begin{equation}\label{eqapp2_10}
\beta=\frac{B}{\ln2} \ln\frac{{B} g_1}{{(\ln2)}\sigma^2\gamma}
-\frac{B}{\ln2}+\frac{\sigma^2\gamma}{ g_1},
\end{equation}
which indicates that $\beta$ is a function of $\gamma$.
Based on (\ref{eqapp2_3}) and (\ref{appendixAeq2}), we have
\begin{equation}\label{eqapp2_4}
\alpha_j=\frac{{(\ln2)}\sigma^2\gamma}{{B}g_j} w^{-1}\left(\frac{g_j \beta }{\sigma^2 \gamma}\right)-1,\quad j=2, \cdots, M.
\end{equation}
where $w^{-1}(x)$ is the inverse function of $w(x)$.
Plugging (\ref{eqapp2_4}) into (\ref{eqapp2_1}), we can obtain
\begin{equation}\label{eqapp2_5}
1+\frac{g_j \bar p_j}{\sigma^2 m_j}=w^{-1}\left(\frac{g_j \beta }{\sigma^2 \gamma}\right), \quad j =2, \cdots, M.
\end{equation}
Since $\alpha_M>\cdots>\alpha_2>\alpha_1=0$, constraints (\ref{KKT2}g) hold with equality for $j=2, \cdots, M$.
Applying (\ref{eqapp2_5}) to (\ref{KKT2}g) yields
\begin{equation}\label{eqapp2_12}
m_j =\frac{{(\ln2)} D_j}{{B} \ln\left( w^{-1}\left(\frac{g_j \beta }{\sigma^2 \gamma}\right)\right)},\quad j =2 , \cdots, M.
\end{equation}
Combining (\ref{eqapp2_5}) and (\ref{eqapp2_12}), we have
\begin{equation}\label{eqapp2_11}
\bar p_j= \frac{{(\ln2)} D_j \sigma^2
\left(w^{-1}\left(\frac{g_j \beta }{\sigma^2 \gamma}\right)-1\right)
}{{B} g_j\ln\left( w^{-1}\left(\frac{g_j \beta }{\sigma^2 \gamma}\right)\right)}, \quad j =2 , \cdots, M,
\end{equation}
where $\beta$ is a function of $\gamma$ from (\ref{eqapp2_10}).
According to (\ref{eqapp2_11}), we find that $\bar p_j$ is a function of $\gamma$ for $j =2 , \cdots, M$.
Then, we show that $\bar p_j$ is decreasing with $\gamma$, $\forall j \geq 2$.
According to (\ref{eqapp2_10}), we have
\begin{equation}\label{eqapp2_14}
\frac{\partial \left(\frac{g_j \beta }{\sigma^2 \gamma}\right)}{\partial \gamma}
=-\frac{B g_j}{(\ln2)\sigma^2 \gamma^2} \ln\frac{{B} g_1}{{(\ln2)}\sigma^2\gamma} <0,
\end{equation}
which implies that $\frac{g_j \beta }{\sigma^2 \gamma}$ is decreasing for $\gamma$.
Moreover, we also can obtain
\begin{equation}\label{eqapp2_15}
\frac{\partial \left(\frac{x-1}{ \ln x }\right)}{\partial x}
= \frac{x \ln x -x +1}
{x\ln^2(x)} >0, \quad \forall x>1.
\end{equation}
According to the chain rule of composition function, we can observe that $\bar p_j$ is decreasing with $\gamma$ from (\ref{eqapp2_11}), (\ref{eqapp2_14}) and (\ref{eqapp2_15}), $j =2 , \cdots, M$.

Owing to the fact that $\beta>0$ implied by (\ref{eqapp_2_15}), constraint (\ref{KKT2}h) holds with equality from (\ref{KKT2}d).
Substituting (\ref{eqapp2_12}) into (\ref{KKT2}h), we have
\begin{equation}\label{eqapp2_6}
m_1= 1- \sum_{j=2}^M \frac{{(\ln2)} D_j}{{B} \ln\left(w^{-1}\left(\frac{g_j \beta }{\sigma^2 \gamma}\right)\right)}.
\end{equation}
Applying $\alpha_1=0$ to (\ref{eqapp2_1}) yields
\begin{equation}\label{eqapp2_7}
\bar p_1=\frac{Bg_1-(\ln2) \sigma^2 \gamma}{(\ln2)g_1\gamma}m_1.
\end{equation}
Substituting (\ref{eqapp2_6}) into (\ref{eqapp2_7}), we have
\begin{equation}\label{eqapp2_13}
\bar p_1=\frac{Bg_1-(\ln2) \sigma^2 \gamma}{(\ln2)g_1\gamma} \left(
1- \sum_{j=2}^M \frac{{(\ln2)} D_j}{{B} \ln\left(w^{-1}\left(\frac{g_j \beta }{\sigma^2 \gamma}\right)\right)}
\right).
\end{equation}
Because both positive-valued function $\frac{Bg_1-(\ln2) \sigma^2 \gamma}{(\ln2)g_1\gamma}$  and $
1- \sum_{j=2}^M \frac{{(\ln2)} D_j}{{B} \ln\left(w^{-1}\left(\frac{g_j \beta }{\sigma^2 \gamma}\right)\right)}
$ are decreasing with $\gamma$, we can prove that $\bar p_1$ is decreasing with $\gamma$.
Since $\gamma>0$ according to (\ref{eqapp2_1}), constraint (\ref{KKT2}i) holds with equality from (\ref{KKT2}e), i.e.,
\begin{equation}\label{eqapp2_16}
 \sum_{j=1}^M \bar p_j=P_{\max}.
\end{equation}
From the above analysis, we prove that the left term of Equation (\ref{eqapp2_16}) is decreasing with $\gamma$, which means that the unique $\gamma$ can be solved from (\ref{eqapp2_16}) by using the bisection method.
Having obtained the value of $\gamma$, the optimal $\pmb {\tilde m }$ and $\pmb {\bar q}$ can be obtained from (\ref{eqapp2_10}), (\ref{eqapp2_12}), (\ref{eqapp2_11}), (\ref{eqapp2_6}) and (\ref{eqapp2_7}).

\section{Proof of Theorem 3}
To show this, if the pair $(\pmb{m},\pmb{p},\pmb{D})$ is feasible in (\ref{max1_1}), then the pair
$(\pmb{m}, \pmb{q})$, where power $q_i=\sum_{j \in {\cal{J}}_i }m_{ij} p_{ij}$, $\forall i \in \cal{N} $,
is feasible in (\ref{max1_2}),
with the same objective value $\sum_{i \in {\cal{N}}} q_{i}=\sum_{i \in {\cal{N}}}\sum_{j \in {\cal{J}}_i }m_{ij} p_{ij}$.
It follows that the optimal value of (\ref{max1_1}) is greater than or equal to the optimal value of (\ref{max1_2}).

Conversely, if $(\pmb{m}, \pmb{q})$ is the optimal solution of (\ref{max1_2}),
we can claim that
\begin{equation}
q_i = \sum_{j \in {\cal{J}}_i } m_{ij} h_{ij}(m_{ij},\pmb{q}_{-i}, D_{ij}), \forall i  \in {\cal{N}}.
\end{equation}
If there exists at least one $n \in {\cal{N}}$ which satisfies $q_n > \sum_{j \in {\cal{J}}_n } m_{nj} h_{nj}(m_{nj},\pmb{q}_{-n}, D_{nj})$.
With $\pmb{m}$ and $\pmb q_{-n}$ fixed, let
$q_n'=\sum_{j \in {\cal{J}}_n } m_{nj} h_{nj}(m_{nj},\pmb{q}_{-n}, D_{nj})$.
Denote the new average power of BSs as $\pmb{q}'=(q_1, \cdots, q_{n-1}, q_n',$ $q_{n+1}, \cdots, q_{N})^T$.
From (\ref{eq4_2}) and (\ref{eq5}),
for all $ k \neq n$, we have
\begin{equation*}
\sum_{l \in {\cal{J}}_k } \!\!m_{kl} h_{kl}(m_{kl},\pmb{q}'_{-k}, D_{kl})\!<\!\!\sum_{l \in {\cal{J}}_k }\! \!m_{kl} h_{kl}(m_{kl},\pmb{q}_{-k}, D_{kl})\!\leq\! q_k.
\end{equation*}
Then, $(\pmb{m}, \pmb{q}')$ is feasible with
\begin{equation}
q_n'+ \sum_{k \in {\cal{N}}\setminus\{n\}} q_k <\sum_{i \in {\cal{N}}} q_i,
\end{equation}
which contradicts the fact that $(\pmb{m, q})$ is the optimal solution.
Thus, $q_i = \sum_{j \in {\cal{J}}_i } m_{ij} h_{ij}($ $m_{ij},\pmb{q}_{-i}, D_{ij}), \forall i  \in {\cal{N}}$.
The pair $(\pmb{m, p, D})$, where $p_{ij}=h_{ij}(m_{ij},\pmb{q}_{-i}, D_{ij}), \forall i  \in {\cal{N}}, \forall j \in {\cal{J}}_i$, is feasible in Problem (\ref{max1_1}) with the same objective value $\sum_{i \in {\cal{N}}}\sum_{j \in {\cal{J}}_i }m_{ij} p_{ij}=\sum_{i \in {\cal{N}}} q_{i}$.
Thus, we conclude that the optimal value of (\ref{max1_1}) is less than or equal to the optimal value of (\ref{max1_2}).
Hence, Problem (\ref{max1_1}) is equivalent to Problem (\ref{max1_2}).

\section{Proof of Theorem 4}
We prove each of the three properties required for standard function below.

Positivity: Through above analysis in $\text{\uppercase\expandafter{\romannumeral5}-B}$,
we have
$v_i(\pmb{q})=\sum_{j \in {\cal{J}}_i }
a_{ij} m_{ij}
(\text e^{b_{ij}/ m_{ij} }-1 )$, where
$m_{ij}={b_{ij}}/ { u^{-1} ( { \lambda_i} /{a_{ij}} )}$
and $\lambda_i$ satisfies $\hat u_i(\lambda_i)=1$.
For all $\pmb{q} \geq \pmb{0}$, $a_{ij}=\frac{\sum_{k \in {\cal{N}}\setminus \{i\}}  q_k g_{kj} +\sigma ^2}
{g_{ij}}>0$, $b_{ij}=\frac {\ln(2)D_{ij}}  {B} >0$, $\forall j \in {\cal{J}}_i$.
Thus, according to (\ref{eq11}), $\lambda_i>0$, $m_{ij}>0$, $\forall j \in {\cal{J}}_i$ and $v_i(\pmb{q})>0$, $\forall i \in {\cal{N}}$, i.e., $\pmb{v}(\pmb{q})>\pmb{0}$.

Monotonicity:
Let power vector $\pmb{q}^{(1)}=(q_1^{(1)}, \cdots, q_N^{(1)})^T$
and $\pmb{q}^{(2)}=(q_1^{(2)}, \cdots, q_N^{(2)})^T$ be such
that $q_k^{(1)} \geq q_k^{(2)}$, $\forall k \in \cal{N}$.
Define
$a_{ij}^{(1)}=({\sum_{k \in {\cal{N}}\setminus \{i\}}  q_k^{(1)} g_{kj} +\sigma ^2})/
{g_{ij}}$ and
$a_{ij}^{(2)}=({\sum_{k \in {\cal{N}}\setminus \{i\}}  q_k^{(2)} g_{kj} +\sigma ^2})/
{g_{ij}}$, $\forall j \in {\cal{J}}_i$.
Obviously, we have $a_{ij}^{(1)} \geq a_{ij}^{(2)}$ , $\forall j \in {\cal{J}}_i$.
We also define ${ \lambda_i^{(1)}}$ and ${ \lambda_i^{(2)}}$ which satisfy
\begin{equation}\label{eqnarray11}
\sum_{j \in {\cal{J}}_i }
\frac{b_{ij}} { u^{-1} \left(\frac { \lambda_i^{(1)}} {a_{ij}^{(1)}} \right)}=1, \quad
\sum_{j \in {\cal{J}}_i }
\frac{b_{ij}} { u^{-1} \left(\frac { \lambda_i^{(2)}} {a_{ij}^{(2)}} \right)}=1.
\end{equation}
Denote the load $m_{ij}^{(1)}={b_{ij}}/ { u^{-1} ( { \lambda_i^{(1)}} /{a_{ij}^{(1)}} )}$
and $m_{ij}^{(2)}={b_{ij}}/ { u^{-1} ( { \lambda_i^{(2)}} /{a_{ij}^{(12)}} )}$,
$\forall j \in {\cal{J}}_i$.
Then, we have
\begin{eqnarray}\label{eqnarray12}
v_i(\pmb{q}^{(1)})
&&\!\!\!\!\!\!\!=\sum_{j \in {\cal{J}}_i }
a_{ij}^{(1)} m_{ij}^{(1)}
(\text e^{b_{ij}/ m_{ij}^{(1)} }-1 )
\nonumber \\
&&\!\!\!\!\!\!\!\geq
\sum_{j \in {\cal{J}}_i }
a_{ij}^{(2)} m_{ij}^{(1)}
( \text e^{b_{ij}/ m_{ij}^{(1)} }-1 )\nonumber \\
&&\!\!\!\!\!\!\!\geq
\sum_{j \in {\cal{J}}_i }
a_{ij}^{(2)} m_{ij}^{(2)}
( \text e^{b_{ij}/ m_{ij}^{(2)} }-1 )=v_i(\pmb{q}^{(2)}).
\end{eqnarray}
The first inequality follows from the fact that $a_{ij}^{(1)} \!\geq\! a_{ij}^{(2)}$, $\forall j \in {\cal{J}}_i$.
The second inequality follows
because $(m_{i (J_{i-1}+1)}^{(2)}, \cdots, m_{i J_{i}}^{(2)})^T$
is the optimal solution of (\ref{max1_3}) with objective function
$\sum_{j \in {\cal{J}}_i }
a_{ij}^{(2)} m_{ij}( e^{b_{ij}/ m_{ij} }-1 )$.
As a result,
${v}_i(\pmb{q}^{(1)}) \geq {v}_i(\pmb{q}^{(2)})$,
$\forall i \in {\cal{N}}$.

Scalability:
Let $\pmb{q} \geq \pmb{0}$ and $\alpha >1$.
Denote load vector $(m_{i (J_{i-1}+1)}, \cdots, m_{i J_{i}})^T$ as
the optimal solution of Problem (\ref{max1_3}).
Let
$a_{ij}'=({\sum_{k \in {\cal{N}}\setminus \{i\}}  \alpha q_k g_{kj} +\sigma ^2}
)/{g_{ij}}$ and $(m_{i (J_{i-1}+1)}', \cdots, m_{i J_{i}}')^T$ the optimal solution
of Problem (\ref{max1_3}) with objective function $\sum_{j \in {\cal{J}}_i }
a_{ij}' m_{ij}( e^{b_{ij}/ m_{ij} }-1 )$ and constraint (\ref{max1_3}b).
Then, we have
\begin{eqnarray}\label{eqnarray13}
\alpha v_i(\pmb{q})
&&\!\!\!\!\!\!\!=\sum_{j \in {\cal{J}}_i }
\alpha a_{ij} m_{ij}
( \text e^{b_{ij}/ m_{ij} }-1 )
\nonumber \\
&&\!\!\!\!\!\!\!>
\sum_{j \in {\cal{J}}_i }
a_{ij}' m_{ij}
(\text  e^{b_{ij}/ m_{ij} }-1 )\nonumber \\
&&\!\!\!\!\!\!\!\geq
\sum_{j \in {\cal{J}}_i }
a_{ij}' m_{ij}'
( \text e^{b_{ij}/ m_{ij}' }-1 )=v_i(\alpha \pmb{q}).
\end{eqnarray}
The first inequality follows from the observation that noise power $\sigma ^2>0$ and $\alpha a_{ij}= ({\sum_{k \in {\cal{N}}\setminus \{i\}}  \alpha q_k g_{kj} +\alpha \sigma ^2}
)/{g_{ij}} > ({\sum_{k \in {\cal{N}}\setminus \{i\}}  \alpha q_k g_{kj} + \sigma ^2}
)/{g_{ij}} = a_{ij}'$.
Meanwhile,
it is obvious that both
$(m_{i (J_{i-1}+1)}, \cdots, m_{i J_{i}})^T$
and
$(m_{i (J_{i-1}+1)}', \cdots,$ $ m_{i J_{i}}')^T$ satisfy the constraint (\ref{max1_3}b).
Similarly to the proof of monotonicity, the second inequality follows
from the definition of $(m_{i (J_{i-1}+1)}', \cdots, m_{i J_{i}}')^T$.
As a result,
$\alpha{v}_i(\pmb{q}) > {v}_i(\alpha \pmb{q})$,
$\forall i \in {\cal{N}}$.

\section{Proof of Corollary 2}
On one hand, if Problem (\ref{max1_2}) is feasible, there exists one pair ($\pmb m, \pmb q$) satisfies (\ref{max1_2}b), (\ref{max1_2}c) and $\pmb q \leq \pmb Q^{\max}$.
Since $v_{i}(\pmb q)$ is the optimal solution of Problem (\ref{max1_3}), we have
\begin{equation}
v_{i} (\pmb q)\leq \sum_{j \in {\cal{J}}_i } m_{ij} h_{ij}(m_{ij},\pmb{q}_{-i}, d_{ij})\leq q_i, \quad\forall i  \in {\cal{N}}.
\end{equation}
Thus, $\pmb Q^{\max}\geq \pmb q \geq \pmb v(\pmb q)$.
Suppose $\pmb v^{(k-1)}(\pmb q)\geq\pmb v^{(k)}(\pmb q)$, then
\begin{equation}
\pmb v^{(k)}(\pmb q)=\pmb v(\pmb v^{(k-1)}(\pmb q)) \geq \pmb v(\pmb v^{(k)}(\pmb q))=\pmb v^{(k+1)}(\pmb q)
\end{equation}
is satisfied by the monotonicity property of Theorem 4.
By using induction method, $\pmb Q^{\max}\geq \pmb v(\pmb q)\geq \pmb v^2(\pmb q))\geq\cdots\geq \pmb v^n(\pmb q)$.
From Corollary 1, the iterative fixed-point method converges, and
$\pmb Q^{\max}\geq \lim_{n\rightarrow\infty}\pmb v^n(\pmb q)=\pmb q^*$.

On the other hand, if there exists $\pmb q^*=\pmb v(\pmb q^*)$ and $\pmb q^* \leq \pmb Q^{\max}$,
the optimal $\pmb m_i^*$ can be obtained by solving Problem $(\ref{max1_3})$ with fixed $\pmb q_{-i}^*$.
Denote $\pmb m^*=((\pmb m^*_1)^T, \cdots, (\pmb m^*_N)^T)^T$.
Obviously, the pair ($\pmb m^*, \pmb q^*$) satisfies all the constraints of Problem (\ref{max1_2}).
Thus, Problem (\ref{max1_2}) is feasible.

\section{Proof of Corollary 3}
According to the definition of $ v_i (\pmb q)$, constraints (\ref{max1_2}b) and  (\ref{max1_2}c) indicate that $v_i (\pmb q) \leq q_i$, $\forall i \in \mathcal N$.
As a result, Problem  (\ref{max1_2}) can be equivalently transformed into
\begin{subequations}\label{max2_3}
\begin{align}
\mathop{\min}_{\pmb{0} < \pmb{q} \leq \pmb{Q}^{\max}}\quad
&\sum_{i \in {\cal{N}}} q_i
\\
\textrm{s.t.}\qquad\:
& v_i (\pmb q) \leq q_i, \quad\forall i \in \mathcal N.
\end{align}
\end{subequations}

To show the equivalence of Problem (\ref{max1_2}) and Problem (\ref{max2_3}), if the pair $(\pmb{m},\pmb{q})$ is feasible in (\ref{max2_3}), then the pair
$(\pmb{q})$ is feasible in (\ref{max1_2}),
with the same objective value.
It follows that the optimal value of (\ref{max1_2}) is greater than or equal to the optimal value of (\ref{max2_3}).
From Theorem 4,
$\pmb{v}(\pmb{q})$ is strictly increasing.
Conversely, assume that $\pmb{q}$ is feasible (\ref{max1_2}).
Denote the optimal solution of Problem (\ref{max1_3}) as $\pmb m_i$ which satisfies constraint (\ref{max1_3}b).
The pair $(\pmb{m},\pmb{q})$, where $\pmb m=(\pmb m_1^T, \cdots, \pmb m_N^T)^T$ is feasible in (\ref{max2_3}) with the same objective value.
Thus, we conclude that the optimal value of (\ref{max1_2}) is less than or equal to the optimal value of (\ref{max2_3}).
As a result, Problem (\ref{max1_2}) is equivalent to Problem (\ref{max2_3}).

From Theorem 4,
$\pmb{v}(\pmb{q})$ is strictly increasing.
Hence, if Program (\ref{max2_3}) is feasible, then for any optimal solution to (\ref{max2_3}),
constraints (\ref{max2_3}b) hold with equality, as otherwise (\ref{max2_3}a) can be improved,
contradicting that the solution is optimal.
As a result, the optimal $\pmb q^*$ of Problem (\ref{max1_2}) satisfies  $\pmb q^*=\pmb v(\pmb q^*)$.
According to Corollary 1, the solution of $\pmb q^*$ is unique.
With fixed $\pmb q_{-i}^*$, the optimal $\pmb m_i^*$ can be obtained by solving Problem $(\ref{max1_3})$.
Since Equation (\ref{eq11}) has unique solution $\lambda_i^*>0$ and $m_{ij}^*=\frac{b_{ij}} { u^{-1} \left(\frac { \lambda_i^*} {a_{ij}} \right)}$, the optimal $\pmb m_i^*$ is also unique.
Thus, the optimal solution ($\pmb m^*, \pmb q^*$) of Problem (\ref{max1_2}) is unique.


\section{Proof of Theorem 6}
The proof is established by showing that when one BS updates its load and power vector by solving Problem (\ref{ratemax1_2_2}), the sum rate of all BSs is non-decreasing.
Let $\pmb{m}=(\pmb{m}_1^T, \cdots, \pmb{m}_N^T)^T$, $\pmb{p}=(\pmb{p}_1^T, \cdots, \pmb{p}_N^T)^T$ and $\pmb{r}=(\pmb{r}_1^T, \cdots, \pmb{r}_N^T)^T$ respectively denote the load, power and rate vectors of all BSs before BS $i$ starts to update its load, power and rate vectors.
Assume that ($\pmb m, \pmb p, \pmb r$) is a feasible solution of Problem (\ref{max1_1}).
Let ($\pmb{\tilde m}_i$, $\pmb {\tilde p}_i$, $\pmb{\tilde r}_i$) denote the updated load, power and rate vectors of BS $i$ with given ($\pmb m_{-i}, \pmb p_{-i}, \pmb r_{-i}$).
According to (\ref{ratemax1_2_2}) and (\ref{ratemax1_2_3}), we can observe that ($\pmb {\tilde m}, \pmb {\tilde p}, \pmb {\tilde r}$) is also a feasible solution of sum rate optimization Problem (\ref{max1_1}).
Then, we have
\begin{subequations}
\begin{align}
\sum_{i \in \mathcal N} \sum_{j\in\mathcal J_i} r_{ij} =&
\sum_{k \in \mathcal N} \sum_{l\in\mathcal J_k} r_{kl}+ \sum_{j \in {\cal{J}}_i }r_{ij} \nonumber\\
&
\leq
\sum_{k \in \mathcal N} \sum_{l\in\mathcal J_k} r_{kl}+ \sum_{j \in {\cal{J}}_i }\tilde r_{ij}
\leq \sum_{i \in \mathcal N} \sum_{j\in\mathcal J_i} \tilde r_{ij}\nonumber,
\end{align}
\end{subequations}
where the first and second equality follow from the fact that ($\pmb{\tilde m}_i$, $\pmb {\tilde p}_i$, $\pmb{\tilde r}_i$) is the load, power and rate vectors of BS $i$ by solving (\ref{ratemax1_2_2}) with given ($\pmb m_{-i}, \pmb p_{-i}, \pmb r_{-i}$). Hence, the DTAPC-RM algorithm must converge.

\bibliographystyle{IEEEtran}
\bibliography{MMM}

\end{document}